\documentclass[11pt]{article}
\usepackage[utf8]{inputenc}
\usepackage[margin=2.5cm]{geometry}
\usepackage{amsmath}
\usepackage{amsfonts}
\usepackage{natbib}
\usepackage[]{algorithm}
\usepackage{algpseudocode}
\usepackage{graphicx}
\usepackage{cases}
\usepackage{comment}
\usepackage{subcaption}
\usepackage{xcolor}
\usepackage{mathtools}
\usepackage{url}

\title{Sequentially guided MCMC proposals for synthetic likelihoods and correlated synthetic likelihoods}
\author{Umberto Picchini$^1$\footnote{picchini@chalmers.se}, Umberto Simola$^2$, Jukka Corander$^2$ $^3$}
\date{\small $^1$ Dept. Mathematical Sciences, Chalmers University of Technology and the University of Gothenburg
Sweden
\\
\small $^2$ Department of Mathematics and Statistics, University of Helsinki, Finland
\\
$^3$ Department of Biostatistics, University of Oslo, Norway}

\begin{document}

\maketitle

\begin{abstract}
Synthetic likelihood (SL) is a strategy for parameter inference when the likelihood function is analytically or computationally intractable. In SL, the likelihood function of the data is replaced by a multivariate Gaussian density over summary statistics of the data. SL requires simulation of many replicate datasets at every parameter value considered by a sampling algorithm, such as Markov chain Monte Carlo (MCMC), making the method computationally-intensive. We propose two strategies to alleviate the computational burden. First, we introduce an algorithm producing a proposal distribution that is sequentially tuned and made conditional to data, thus it rapidly \textit{guides} the proposed parameters towards high posterior density regions. In our experiments, a small number of iterations of our algorithm is enough to rapidly locate high density regions, which we use to initialize one or several chains that make use of off-the-shelf adaptive MCMC methods. Our ``guided'' approach can also be potentially used with MCMC samplers for approximate Bayesian computation (ABC).
Second, we exploit strategies borrowed from the correlated pseudo-marginal MCMC literature, to improve the chains mixing in a SL framework. 
Moreover, our methods enable inference for challenging case studies, when the posterior is multimodal and when the chain is initialised in low posterior probability regions of the parameter space, where standard samplers failed. 
To illustrate the advantages stemming from our framework we consider five benchmark examples, including estimation of parameters for a cosmological model and a stochastic model with highly non-Gaussian summary statistics. 
\end{abstract}

\textbf{Keywords:} Bayesian inference; cosmological parameters; intractable likelihoods; likelihood-free.

\section{Introduction}

Synthetic likelihood (SL) is a methodology for parameter inference in stochastic models that do not admit a computationally tractable likelihood function. That is, similarly to approximate Bayesian computation (ABC, \citealp{sisson2018handbook}), SL only requires the ability to generate synthetic datasets from a model simulator, and statistically relevant summary statistics of the data that capture parameter-dependent variation in an adequate manner. The price to pay for its flexibility is that SL can be computationally very intensive, since it is typically embedded into a Markov chain Monte Carlo (MCMC) framework, requiring the simulation of multiple (often hundreds or thousands) synthetic datasets at each proposed parameter. The goal of our work is twofold: (i) we introduce an algorithm that sequentially produces a proposal sampler that is made conditional to data and rapidly enables the identification of high-posterior-density regions, where to initialize MCMC chains using off-the-shelf methods; (ii) we introduce a way to increase the chains mixing, by tweaking methods that have been recently proposed in the correlated particle filters literature. Hence both strategies aim at reducing the computational cost to perform Bayesian inference via SL. We show that our approaches facilitate sampling when the chains are initialised at parameter values in regions of low posterior probability, a case where SL often struggles, see the case studies in sections \ref{sec:astro-simul} and \ref{sec:boom-bust} where the Bayesian synthetic likelihoods (BSL) of \cite{price2018bayesian} fails when using the adaptive MCMC proposal of \cite{haario2001adaptive}. For the case study in section \ref{sec:boom-bust}, having strongly non-Gaussian summary statistics, we show that even a BSL version robustified to non-Gaussian summaries fails to explore the posterior surface when initialized at challenging locations, while our proposal sampler is able to quickly converge towards the high-density region. Our proposal sampler can be beneficial with multimodal targets, to inform the researcher of the existence of multiple modes using a small number of iterations, see section \ref{sec:mixture}. 
In addition, we inform the reader that for challenging problems where it is difficult to locate appropriate starting parameters, an alternative to our method is Bayesian optimization, which  can be efficiently used for kickstarting SL-based posterior sampling \citep{gutmann2016bayesian}, and is facilitated by the open-source \texttt{ELFI} software \citep{lintusaari2018elfi}.

SL is described in detail in Section \ref{sec:sl}, but here we first review its features with relevant references to the literature. SL was first proposed in \cite{wood2010statistical} to produce inference for parameters $\theta$ of simulator-based models with an intractable likelihood. SL replaces the analytically intractable data likelihood $p(y|\theta)$ for observed data $y$ with the joint density of a set of summary statistics of the data $s:=T(y)$. Here $T(\cdot)$ is a function of the data that has to be specified by the analyst and that can be evaluated for input $y$, or simulated data $y^*$. The SL approach is characterized by the assumption  that $s$ has a multivariate normal distribution $s\sim \mathcal{N}(\mu_\theta,\Sigma_\theta)$ with unknown mean $\mu_\theta$ and covariance matrix $\Sigma_\theta$. These can be estimated via Monte Carlo simulations of size $M$ to obtain estimators $\hat{\mu}_{M,\theta}$, $\hat{\Sigma}_{M,\theta}$.
The resulting multivariate Gaussian likelihood ${p}_M(s|\theta)\equiv \mathcal{N}(\hat{\mu}_{M,\theta},\hat{\Sigma}_{M,\theta})$ can then be numerically maximised with respect to $\theta$, to return an approximate maximimum likelihood estimator \citep{wood2010statistical}. It can also be plugged into a Metropolis-Hastings algorithm with flat priors \citep{wood2010statistical}, so that MCMC is used as a workhorse to sample from a posterior $\pi_M(\theta|s)$ to ultimately return the posterior mode, and hence a maximum likelihood estimator (a purely Bayesian approach is described below). The introduction of data summaries in the inference has been shown to cope well with chaotic models, where the likelihood would otherwise be difficult to optimize and the corresponding posterior surface may be difficult to explore. More generally, SL is a tool for likelihood-free inference, just like the ABC framework (see reviews \citealp{sisson2011likelihood,karabatsos2018approximate}), where the latter can be seen as a nonparametric methodology, while SL uses a parametric distributional assumption on $s$. SL has found applications in e.g. ecology \citep{wood2010statistical}, epidemiology \citep{engblom2019bayesian,dehideniya2019synthetic}, mixed-effects modeling of tumor growth \citep{picchini2019bayesian}. For a recent generalization of the SL family of inference methods using statistical classifiers to directly target estimation of the posterior density, see \cite{thomas2016likelihood} and \cite{kokko2019pylfire}. 

While ABC is more general than SL, it can sometimes be difficult to tune and it typically suffers from the ``curse of dimensionality'' when the size of $s$ increases, due to its nonparametric nature. 
On the other hand, the Gaussianity assumption concerning the summary statistics is the main limitation of SL. At the same time, due to its parametric nature, SL has been shown to perform satisfactorily on problems where $\dim(s)$ is large relative to $\dim(\theta)$ \citep{ong2018likelihood}. \cite{price2018bayesian} framed SL within a pseudo-marginal algorithm for Bayesian inference \citep{andrieu2009pseudo}  and named the method Bayesian SL (BSL). They showed that when $s$ is truly Gaussian, BSL produces MCMC samples from $\pi(\theta|s)$, not depending on the specific choice of $M$. However, in practice, the inference algorithm does depend on the specific choice of $M$, since this value affects the chains mixing. 
Unless the underlying computer model is trivial, producing the $M$ datasets for each $\theta$ can be a serious computational bottleneck. 

In this work we design a strategy producing a proposal sampler $g(\cdot|s)$ that is conditional to summary statistics of the data, by exploiting the Gaussian assumption for the summary statistics in (B)SL. We call this a \textit{guided sampler}, as it proposes conditionally to data. Moreover, our guided sampler is sequentially built, and we find that our ``sequentially Adapted and guided proposal for SL'' (named ASL) is easy to construct and adds essentially no overhead, since it exploits quantities that are anyway computed in SL. We stress the importance of rapid convergence to the bulk of the posterior, as while SL may require a large $M$ to get started, once it has approached high posterior probability regions $M$ can be reduced substantially (in section \ref{sec:g-and-k} we are forced to start with $M=1,000$ and after a few iterations we can revert to $M=10$ or 50). Later we briefly discuss how the proposal sampler can also be used in an ABC-MCMC algorithm. We emphasize that our algorithm should be used to rapidly identify the high density region of the posterior, and there initialize other algorithms to produce the actual inference (e.g. using some of the several available adaptive MCMC samplers). We discuss this aspect and suggest possibilities afterwards. In section \ref{sec:mixture} we show how ASL can be useful with multimodal targets.
 
 Furthermore, we correlate log-synthetic likelihoods using a ``blockwise'' strategy, borrowed from relatively recent advances in pseudo-marginal MCMC literature. This is shown to considerably improve mixing of the chains generated via SL, while not introducing correlation can lead to unsatisfactory simulations when using starting parameter values residing relatively far from the representative ones. 
 Finally, we wish to inform the reader of a further possibility to initialize SL algorithms, besides our sequentially adaptive proposal sampler: Bayesian optimization via the BOLFI method \citep{gutmann2016bayesian} available in \texttt{ELFI} \citep{lintusaari2018elfi}, the engine for likelihood-free inference. 
 
 Our paper is structured as follows: in Section \ref{sec:sl} we introduce the synthetic likelihood approach. In Section \ref{sec:asl} we construct the adaptive proposal distribution via ASL and in section \ref{sec:corrsynlik} we construct correlated synthetic likelihoods. In Section \ref{sec:bolfi} we discuss using BOLFI and \texttt{ELFI} as an option for SL inference. In Section \ref{sec:simulation-studies} we discuss four benchmarking simulation studies and a fifth one is in Supplementary Material. Code can be found at \url{https://github.com/umbertopicchini/ASL}.

\section{Synthetic likelihood}\label{sec:sl}

We briefly summarize the synthetic likelihood (SL) method as proposed in \cite{wood2010statistical} and in a Bayesian context in \cite{price2018bayesian} (the latter is detailed in Supplementary Material). The main goal is to produce Bayesian inference for $\theta$, by sampling from (an approximation to) the posterior $\pi(\theta|s)\propto \tilde{p}(s|\theta)\pi(\theta)$ using MCMC, where $\tilde{p}(s|\theta)$ is the density underlying the true (unknown) distribution of $s$.
\cite{wood2010statistical} proposes a parametric approximation to $\tilde{p}(s|\theta)$, placing the rather strong assumption that $s\sim \mathcal{N}(\mu_\theta,\Sigma_\theta)$. The reason for this assumption is that estimators for the unknown mean and covariance of the summaries, $\mu_\theta$ and $\Sigma_\theta$ respectively, can be obtained straightforwardly via simulation, as described below. As obvious from the notation used, $\mu_\theta$ and $\Sigma_\theta$ depend on the unknown finite-dimensional vector parameter $\theta$, and these are estimated by simulating independently $M$ datasets from the assumed data-generating model, conditionally on $\theta$. 
We denote the synthetic datasets simulated from the assumed model run at a given $\theta^*$ with $y_1^{*},...,y_M^{*}$. These are such that $\dim(y^*_m)=\dim(y)$ ($m=1,...,M$), with $y$ denoting observed data, and therefore $s\equiv T(y)$. For each dataset it is possible to construct the corresponding (possibly vector valued) summary $s_m^*:=T(y^*_m)$, with $\dim(s_m^{*})=\dim(s)$. These simulated summaries are used to construct the following  estimators:
\begin{equation}
\hat{\mu}_{M,\theta^*} = \frac{1}{M}\sum_{m=1}^{M}s_m^{*},\qquad  
\hat{\Sigma}_{M,\theta^*} = \frac{1}{M-1}\sum_{m=1}^{M}(s_m^{*}-\hat{\mu}_{\theta^*})(s_m^{*}-\hat{\mu}_{\theta^*})'\label{eq:sl-moments},
\end{equation}
with $'$ denoting transposition. By defining ${p}_M(s|\theta)\equiv \mathcal{N}(\hat{\mu}_{M,\theta},\hat{\Sigma}_{M,\theta})$, the SL procedure in \cite{wood2010statistical} samples from the posterior $\pi_M(\theta|s)\propto {p}_M(s|\theta)\pi(\theta)$, see Algorithm \ref{alg:synlikMCMC}. 
A slight modification of the original approach in \cite{wood2010statistical} leads to the ``Bayesian synthetic likelihood'' (BSL) algorithm of \cite{price2018bayesian}, which samples from $\pi(\theta|s)$ when $s$ is truly Gaussian, by introducing an unbiased approximation to a Gaussian likelihood. Besides this, the BSL is the same as Algorithm \ref{alg:synlikMCMC}. See the Supplementary Material for details about BSL. All our numerical experiments use the BSL formulation of the inference problem. Notice when $M$ is too small or $\theta^*$ is implausible, the estimated covariance may mis-behave, e.g. may be not positive-definite: in such case, we attempt a ``modified Cholesky factorization'' of $\hat{\Sigma}_{M,\theta^*}$, such as the one in \cite{cheng1998modified} (we used the Matlab implementation in \citealp{modchol}), or we tried to find a ``nearest symmetric-positive-definite matrix'' \citep{higham1988computing}, using the function by \cite{derrico}.

When the simulator generating the $M$ synthetic datasets is computationally demanding, Algorithm \ref{alg:synlikMCMC} is computer intensive, as it generally needs to be run for a number of iterations $R$ in the order of thousands. The problem is exacerbated by the possibly poor mixing of the resulting chain. The most obvious way to alleviate the problem is to reduce the variance of the estimated likelihoods, by increasing $M$, but of course this makes the algorithm computationally more intensive. A further problem occurs when the initial $\theta^*$ lies far away in the tails of the posterior. This may cause numerical problems when the initial $\hat{\Sigma}_{M,\theta^*}$ is ill-conditioned, possibly requiring a very large $M$ to get the MCMC started, and hence it is desirable to have the chains approach the bulk of the posterior as rapidly as possible. 

In the following we propose two strategies aiming at keeping the mixing rate of a MCMC, produced either by SL or BSL, at acceptable levels and also to ease convergence of the chains to the regions of high posterior density. The first strategy results in designing a specific proposal distribution $g(\cdot)$ for use in MCMC via synthetic likelihood: this is a ``sequentially Adapted and guided proposal for Synthetic Likelihoods'' (shorty ASL) and is described in section \ref{sec:asl}. The second strategy reduces the variability in the Metropolis-Hastings ratio $\alpha$ by correlating successive pairs of synthetic likelihoods: this results in ``correlated synthetic likelihoods'' (CSL) described in section \ref{sec:corrsynlik}.

\begin{algorithm}
\small
\caption{Synthetic likelihoods MCMC}
\begin{algorithmic}
\State   \textbf{Input:} positive integers $M,R$. Observed summaries $s$. Fix starting value $\theta^*$ or generate it from the
prior $\pi(\theta)$. Set $\theta_1:=\theta^*$. Define a proposal  $g(\theta'|\theta)$. Set $r:=1$. \\
\textbf{Output:} $R$ correlated samples from $\pi_M(\theta|s)$. 
\State 1. Conditionally on $\theta^*$ generate independently from the model $M$
summaries $s^{* 1},...,s^{* M}$, compute $\hat{\mu}_{M,\theta^*}$, $\hat{\Sigma}_{M,\theta^*}$ from \eqref{eq:sl-moments}  and
${p}_M(s|\theta^*)\equiv \mathcal{N}(\hat{\mu}_{M,\theta^*},\hat{\Sigma}_{M,\theta^*})$.
\State 2. Generate $\theta^{\#}\sim
g(\theta^{\#}|\theta^*)$. Conditionally on $\theta^\#$ generate
independently $s^{\# 1},...,s^{\# M}$, compute $\hat{\mu}_{M,\theta^\#}$, $\hat{\Sigma}_{M,\theta^\#}$,  and ${p}_M(s|\theta^\#)$. 
\State 3. Generate a uniform random draw $u\sim U(0,1)$, and calculate
the acceptance probability 
\begin{eqnarray*}
\alpha=\min\biggl[1,\frac{{p}_M(s|\theta^\#)}{{p}_M(s|\theta^*)}
\times  \frac{g(\theta^*|\theta^{\#})}{g(\theta^{\#}|\theta^{*})} \times \frac{\pi(\theta^{\#})}{\pi(\theta^*)} \biggr]. 
\end{eqnarray*}
If $u>\alpha$, set $\theta_{r+1}:=\theta_{r}$ otherwise set $\theta_{r+1}:=\theta^{\#}$, $\theta^*:=\theta^{\#}$ and ${p}_M(s|\theta^*):={p}_M(s|\theta^\#)$. Set $r:=r+1$ and go to step 4. 
\State 4. Repeat steps 2--3 as long as $r\leq R$.
\end{algorithmic}
\label{alg:synlikMCMC}
\end{algorithm} 

\section{Guided and sequentially tuned proposals for synthetic likelihoods}\label{sec:asl}

In section \ref{sec:inizialization} we illustrate the main ideas of our ASL method. In section \ref{sec:sequential-kernel} we specialize ASL so that we instead obtain a sequence of proposal distributions $\{g_t\}_{t=1}^T$, as detailed in Algorithm \ref{alg:adaptiveSL}. What we now introduce in section \ref{sec:inizialization} will also initialize the ASL method, i.e. provide an initial $g_0$.

\subsection{Main idea and initialization}\label{sec:inizialization}

Suppose $\theta_n^*$ is a posterior draw generated by some SL procedure (i.e. the standard method from \citealp{wood2010statistical} or the BSL one from \citealp{price2018bayesian}) at iteration $n$, e.g. $\theta_n^*\sim \pi_M(\theta|s)$. Then denote with $\{s^{*1}_n,...,s^{*M}_n\}$ a set of $M$ summaries simulated independently from the computer model, conditionally on the same $\theta_n^*$, and define $\bar{s}_n^*=\sum_{m=1}^M s^{*m}_n/M$. By the central limit theorem, for $M$ sufficiently large $\bar{s}_n$ has an approximately Gaussian distribution.
Suppose we have at disposal $N$ pairs $\{\theta^*_n,\bar{s}^*_n\}_{n=1}^N$.
We set $d_\theta=\dim(\theta)$ and $d_s=\dim(s)$, then $(\theta^*_n,\bar{s}^*_n)$ is a vector having length $d=d_\theta+d_s$. Assume for a moment that the joint vector $(\theta^*_n,\bar{s}^*_n)$ is a $d$-dimensional Gaussian-distributed vector, with $(\theta^*_n,\bar{s}^*_n)\sim \mathcal{N}_d(m,S)$. We stress that this assumption is made merely to construct a proposal sampler, and does not extend to the actual distribution of $(\theta,s)$.
We set a $d$-dimensional mean vector $m\equiv(m_\theta,m_s)$ and the $d\times d$ covariance matrix  
\[
{S}\equiv \left[\begin{array}{cc}
{S}_{\theta} & {S}_{\theta s}   \\
{S}_{s\theta} & {S}_{s} \\
\end{array}\right],
\]
where ${S}_{\theta}$ is $d_\theta\times d_\theta$, ${S}_{s}$ is $d_s\times d_s$, ${S}_{\theta s}$ is $d_\theta \times d_s$ and of course ${S}_{s\theta}\equiv {S}_{\theta s}'$ is $d_s \times d_\theta$.
We estimate $m$ and $S$ using the $N$ available draws. That is, define $x_n:=(\theta^*_n,\bar{s}^*_n)$ then, same as in \eqref{eq:sl-moments}, we have
\begin{equation}
\hat{m} = \frac{1}{N}\sum_{n=1}^{N}x_n,\qquad  
\hat{S} = \frac{1}{N-1}\sum_{n=1}^{N}(x_n-\hat{m})(x_n-\hat{m})'.\label{eq:m-S}
\end{equation}
Once $\hat{m}$ and $\hat{S}$ are obtained, it is possible to extract the corresponding entries $(\hat{m}_\theta,\hat{m}_s)$ and $\hat{S}_\theta$, $\hat{S}_s$, $\hat{S}_{s\theta}$, $\hat{S}_{\theta s}$. We can now use well known formulae for conditionals of a multivariate Gaussian distribution, to obtain a proposal distribution (with a slight abuse of notation) $g(\theta|s)\equiv \mathcal{N}(\hat{m}_{\theta|s},\hat{S}_{\theta|s})$, with 
\begin{align}
    \hat{m}_{\theta|s} &= \hat{m}_{\theta}+\hat{S}_{\theta s}(\hat{S}_s)^{-1}(s-\hat{m}_s) \label{eq:mu-conditional}\\
    \hat{S}_{\theta|s} &= \hat{S}_{\theta}-\hat{S}_{\theta s}(\hat{S}_{s})^{-1}\hat{S}_{s \theta}.\label{eq:cov-conditional}
\end{align}
Hence a new proposal $\theta^*$ can be generated as $\theta^*\sim g(\theta|s)$, and is thus ``guided'' by the summaries of the data  $s$, and gets updated as new posterior draws become available, as further described below. Therefore, this ``guided proposal'' $g(\theta|s)$ can be employed in place of $g(\theta'|\theta)$ into Algorithm \ref{alg:synlikMCMC}, even though we only use this proposal for a limited number of iterations, as clarified below.
Clearly the proposal function $g(\theta|s)$ is independent of the last accepted value of $\theta$, hence it is an ``independence sampler'' \citep{robert2004monte}, except that its mean and covariance matrix are not kept constant. 

The approach outlined so far is essentially step 3 in Algorithm \ref{alg:adaptiveSL}, and together with the sequential tuning in section \ref{sec:sequential-kernel}, allows for a rapid convergence of the chain towards the high posterior density region. However, this approach does not promote tails exploration. This is not really an issue, as we can let an MCMC incorporating our guided proposal sampler run for a small number of iterations (say 50 iterations, even if we use more for pictorial reasons), where the chain displays a high acceptance rate, and this is useful to collect many accepted draws that we can use to initialize other standard samplers enjoying proven ergodic properties, as detailed in next section \ref{sec:sequential-kernel}.
Moreover, the next section also illustrates a sampler based on the multivariate Student's distribution. 

\subsection{Sequential approach}\label{sec:sequential-kernel}

The construction outlined above is only the first step of our guided adaptive sampler for synthetic likelihoods (ASL) methodology, and we now detail it to ease the actual implementation in a sequential way. We define a sequence of $T+1$ ``rounds'' over which $T+1$ kernels $\{g_t\}_{t=0}^T$ are sequentially constructed. In the first round ($t=0$), we  construct $g_0$ using the output of $K\gg N$ MCMC iterations, obtained using e.g. a Gaussian random walk. We may consider $K$ as burnin iterations. Once \eqref{eq:m-S}--\eqref{eq:mu-conditional}--\eqref{eq:cov-conditional} are computed using the output $\{\theta^*_k,\bar{s}^*_k\}_{k=1}^K$ of the burnin iterations, we obtain the first adaptive distribution denoted $g_0(\theta|s)$ as illustrated in section \ref{sec:inizialization}. We store the draws as $\mathcal{D}:=\{\theta^*_k,\bar{s}^*_k\}_{k=1}^K$ and then employ $g_0$ as a proposal density in further $N$ MCMC iterations, after which we perform the following steps: (i) we collect the newly obtained batch of $N$ pairs $\{\theta^*_n,\bar{s}^*_n\}_{n=1}^N$ (where, again, $\theta^*_n\sim \pi_M(\theta|s)$ and $\bar{s}^*_n$ is the sample mean of the \textit{already accepted} simulated summaries generated conditionally to $\theta^*_n$) and add it to the previously obtained ones as $\mathcal{D}:=\mathcal{D}\cup \{\theta^*_n,\bar{s}^*_n\}_{n=1}^N$. Then (ii) similarly to \eqref{eq:m-S} we compute the sample mean $\hat{m}^{0:1}=(\hat{m}^{0:1}_\theta,\hat{m}^{0:1}_s)$ and corresponding covariance $\hat{S}^{0:1}$, except that here $\hat{m}^{0:1}$ and $\hat{S}^{0:1}$ use the $K+N$ pairs in $\mathcal{D}$. (iii)
Update \eqref{eq:mu-conditional}--\eqref{eq:cov-conditional} to $\hat{m}_{\theta|s}^{0:1}$ and $\hat{S}_{\theta|s}^{0:1}$, and obtain $g_1(\theta|s)$. (iv) Use $g_1(\theta|s)$ for further $N$ MCMC moves, stack the new draws into $\mathcal{D}:=\mathcal{D}\cup \{\theta^*_n,\bar{s}^*_n\}_{n=1}^N$, and using the $K+2N$ pairs in $\mathcal{D}$ proceed as before to obtain $g_2$, and so on until the last batch of $N$ iterations generated using $g_T$ is obtained. 

From the procedure we have just illustrated, the sequence of Gaussian kernels has $g_t= g_t(\theta|s)\equiv \mathcal{N}(\hat{m}_{\theta|s}^{0:t},\hat{S}_{\theta|s}^{0:t})$, with $\hat{m}_{\theta|s}^{0:t}$ and $\hat{S}_{\theta|s}^{0:t}$ the conditional mean and covariance matrix given by
\begin{align}
    \hat{m}_{\theta|s}^{0:t} &= \hat{m}_{\theta}^{0:t}+\hat{S}_{\theta s}^{0:t}(\hat{S}_s^{0:t})^{-1}(s-\hat{m}_s^{0:t}) \label{eq:mu-conditional-general}\\
    \hat{S}_{\theta|s}^{0:t} &= \hat{S}_{\theta}^{0:t}-\hat{S}_{\theta s}^{0:t}(\hat{S}_{s}^{0:t})^{-1}\hat{S}_{s \theta}^{0:t}.\label{eq:cov-conditional-general}
\end{align}
The proposal function $g_t$ uses all available present and past information, as these are obtained using the most recent version of $\mathcal{D}$, which contains information from the previous $t-1$ rounds in addition to the latest batch of $N$ draws. 
Compared to a standard Metropolis random walk, the additional computational effort to implement our method is negligible, as it uses trivial matrix algebra applied on quantities obtained as a by-product of the SL procedure, namely the several pairs $\{\theta^*_n,\bar{s}^*_n\}$. Notice \eqref{eq:mu-conditional-general}-\eqref{eq:cov-conditional-general} reduce to $\hat{m}_{\theta|s}^{0:t} \equiv \hat{m}_{\theta}^{0:t}$ and $\hat{S}_{\theta|s}^{0:t} \equiv \hat{S}_{\theta}^{0:t}$ respectively as soon as $\hat{m}_s^{0:t}=s$. The latter condition means that the chain is close to the bulk of the posterior and accepted parameters simulate summaries distributed around the observed $s$. Therefore, when the chain is far from its target, the additional terms in \eqref{eq:mu-conditional-general}-\eqref{eq:cov-conditional-general} can help guide the proposals thanks to an explicit conditioning to data.

An alternative to Gaussian proposals are multivariate Student's proposals. We build on the result found in \cite{ding2016conditional} allowing us to write $\theta^*_n\sim{g}_t(\theta|s)$, and here ${g}_t(\theta|s)$ is a multivariate Student's distribution with $\nu$ degrees of freedom, and in this case $\theta^*_n$ can be simulated using
\begin{equation}
\theta^*_n = \hat{m}_{\theta|s}^{0:t} + \biggl(\sqrt{\frac{\nu+\delta_n}{\nu+d_s}}(\hat{S}_{\theta|s}^{0:t})^{1/2}\biggr)\biggl(Z_n/\sqrt{\frac{\chi^2_{\nu+d_s}}{\nu+d_s}}\biggr)\label{eq:student-prop}
\end{equation}
with $\chi^2_{\nu+d_s}$ an independent draw from a Chi-squared distribution with $\nu+d_s$ degrees of freedom, $\delta_n=(s-\hat{m}_s^{0:t})(\hat{S}_s^{0:t})^{-1}(s-\hat{m}_s^{0:t})'$ and $Z_n$ a $d_\theta$-dimensional standard multivariate Gaussian vector that we simulate at each iteration $n$ and is independent of $\chi^2_{\nu+d_s}/(\nu+d_s)$. For simplicity, in the following we do not make distinction between the Gaussian and the Student's proposals, and the user can choose any of the two, as they are anyway obtained from the same building-blocks \eqref{eq:m-S}--\eqref{eq:cov-conditional-general}.

As customary in Metropolis-Hastings, when a proposal is rejected at a generic iteration $n$, the last accepted pair should be stored as $(\theta_n,\bar{s}_n)$. However, should the rejection rate be high (notice we have never incurred into such situation when sampling via ASL), the covariance $\hat{S}_{\theta s}^{0:t}$ would be computed on many identical repetitions of the same $(\theta,\bar{s})$-vectors, this causing numerical instabilities. Therefore, anytime a rejection takes place, we can perform the following when storing the output of the $n$-th MCMC iteration:

\noindent
\begin{quote}
if proposal $\theta^\#\sim g(\theta|s)$ has been rejected at iteration $n$: resample independently $M$ times with replacement from the last accepted set of summaries $(s^{*1},...,s^{*M})$ (produced from the last accepted $\theta^*$), to obtain the bootstrapped set $(\tilde{s}^{*1},...,\tilde{s}^{*M})$. We use the latter set to compute $\bar{\tilde{s}}^*=\sum_{m=1}^M \tilde{s}^{*m}/M$. Hence, at iteration $n$ (and only when proposal $\theta^\#\sim g(\theta|s)$ is rejected) we store $\mathcal{D}:=\mathcal{D}\cup \{\theta^*_n,\bar{\tilde{s}}^*_n\}$, instead of $\mathcal{D}:=\mathcal{D}\cup \{\theta^*_n,\bar{{s}}^*_n\}$.
\end{quote}

This way, the averaged summaries stored in set $\mathcal{D}$ still consist of averages of accepted summaries (as usual), with the benefit that when the acceptance rate is low (which anyway never occurred to us with ASL) $\hat{S}_{\theta s}^{0:t}$ is computed on a set $\mathcal{D}$ that has more varied information, thanks to resampling. This consideration is expressed in step 5 of Algorithm \ref{alg:adaptiveSL}. Algorithm  \ref{alg:adaptiveSL} constructs the sequence  $\{g_t(\theta|s)\}_{t=1}^T$ for a SL procedure, and this constitutes our ASL approach. 
\begin{algorithm}
\small
\caption{ASL: synthetic likelihoods with a sequentially adapted and guided proposal}
\begin{algorithmic}[1]
\State   \textbf{Input:} $K$ pairs $\{\theta^*_k,\bar{s}^*_k\}_{k=1}^K$ from burnin. Positive integers $N$ and $T$. Initialize $\mathcal{D}:=\{\theta^*_k,\bar{s}^*_k\}_{k=1}^K$.
\State   \textbf{Output:} $\theta_1,...,\theta_T$. Then $\theta_T$ should be used as starting point for another adaptive MCMC algorithm.
\State Construct the proposal density $g_0$ using $\{\theta^*_k,\bar{s}^*_k\}_{k=1}^K$ and \eqref{eq:m-S}--\eqref{eq:mu-conditional}--\eqref{eq:cov-conditional} (and optionally propose from \eqref{eq:student-prop}). Set $\theta_0:=\theta^*_K$.
\For{$t=1:T$}
\State Starting at $\theta_{t-1}$ run $N$ MCMC iterations (SL or BSL) using $g_{t-1}$, producing $\{\theta^*_n,\bar{s}^*_n\}_{n=1}^N$. If the current proposal has been rejected at iteration $n$, the $\bar{s}^*_n$ may instead be computed as $\bar{\tilde{s}}_n^*$ (see main text).
\State Form $\mathcal{D}:=\mathcal{D}\cup\{\theta^*_n,\bar{s}^*_n\}_{n=1}^N$, compute $(\hat{m}^{0:t},\hat{S}^{0:t})$ on $\mathcal{D}$, update $(\hat{m}_{\theta|s}^{0:t},\hat{S}_{\theta|s}^{0:t})$ to construct $g_{t}$.
\State Set $\theta_t:=\theta^*_N$.
\EndFor 
\State Return $\theta_1,...,\theta_T$ to be provided as input to another adaptive MCMC algorithm for BSL or CSL.
\end{algorithmic}
\label{alg:adaptiveSL}
\end{algorithm} 
An advantage of ASL is that it is self-adapting. A disadvantage is that, since the adaptation results into an independence sampler, it does not explore a neighbourhood of the last accepted draw, and newly accepted $N$ draws obtained at stage $t$ might not necessarily produce a rapid change into mean and covariance  for the proposal function $g_{t+1}$ (should a rapid change actually be required for optimal exploration of the parameter space).  This is why in our applications we always use $N=1$. That is, the proposal distribution is updated at each iteration by immediately incorporating information provided by the last accepted draw. 

As clear from the output of Algorithm \ref{alg:adaptiveSL}, we recommend to use $T$ iterations of ASL to return i) $\theta_T$ which is then used as starting parameter value for a run of BSL (or CSL see section \ref{sec:corrsynlik}) together with a standard MCMC proposal sampler; and ii) the sequence $\theta_1,...,\theta_T$ (notice this excludes the initial burnin of $K$ iterations) of which we compute the sample covariance matrix, and the latter is used to initiate the adaptive MCMC algorithm of \cite{haario2001adaptive}, this one having proven ergodic properties (see the Supplementary Material for details on how this is performed). The above means that the inference results we report are based on draws using \cite{haario2001adaptive} (thanks to the useful initialization via ASL). However, after the $T$ ASL iterations, besides \cite{haario2001adaptive} other adaptive MCMC algorithms with proven ergodic properties could be used: possibilities are e.g.   \cite{andrieu2008tutorial} or \cite{vihola2012robust}. Moreover, an interesting use of ASL arise with multimodal targets: if several chains are run in parallel and are initialised at different parameter values, the nature of ASL to rapidly ``jump'' to high density regions can point the researcher to the existence of multiple modes within few iterations of ASL (this is illustrated in section \ref{sec:mixture}). 

In our experiments we use a relatively small number of burnin iterations $K$ (say $K=200$ or 300), and when ASL is started we immediately observe a large ``jump'' towards the posterior mode.  Importantly, rapid convergence via ASL also helps reducing the computational effort by re-tuning $M$: in fact, while a large value of $M$ can be necessary when setting $\theta_0$ in tail regions of the posterior, once the chain has converged towards the bulk of the posterior it is possible to reduce $M$ substantially. See the g-and-k example in section \ref{sec:g-and-k}, where it is necessary to start with $M=1,000$, and after using ASL for a few iterations we can revert to $M=10$ or 50. 

Our ASL strategy is inspired by the sequential neuronal likelihood approach found in \cite{papamakarios2018sequential}. In \cite{papamakarios2018sequential} $N$ MCMC draws obtained in each of $T$ stages sequentially approximate the likelihood function for models having an intractable likelihood, whose approximation at stage $t$ is obtained by training a neuronal network (NN) on the MCMC output obtained at stage $t-1$. Their approach is more general (and it is aimed at approximating the likelihood, not the MCMC proposal), but has the disadvantage of requiring the construction of a NN, and then the NN hyperparameters must be tuned at every stage $t$, which of course requires domain knowledge and computational resources. Our approach is framed specifically for inference via synthetic likelihoods, which is a limitation \textit{per-se}, but it is completely self-tuning, with the possible exception of the burnin iterations where an initial covariance matrix must be provided by the user, though this is a minor issue when the number of parameters is limited. 

Notice, a possible interesting application of our guided sampler could be envisioned with ABC-MCMC algorithms \citep{marjoram2003markov}. Even though ABC-MCMC is typically run by simulating a single vector of summary statistics at a given $\theta$ (though it is also possible to consider pseudo-marginal versions, as in \citealp{picchini2019stratified}), nothing prevents to run ASL for a few iterations in an ABC-MCMC context, by simulating multiple summaries at each $\theta$ as in SL, and then revert to simulating a single summary vector once the chain has reached the bulk of the ABC posterior.

\section{Correlated synthetic likelihood}\label{sec:corrsynlik}

Following the success of the pseudo-marginal method (PM), returning exact Bayesian inference whenever a non-negative and unbiased estimate of an intractable likelihood is available (\citealp{beaumont2003estimation}, \citealp{andrieu2009pseudo}, \citealp{andrieu2010particle}), there has been much research aimed at increasing the efficiency of particle filters (or sequential Monte Carlo) for Bayesian inference in state-space models, see \cite{schon2018probabilistic} for an approachable review. A recent important addition to PM methodology, improving the acceptance rate in Metropolis-Hastings algorithms, considers inducing some correlation between the likelihoods appearing in the Metropolis-Hastings ratio. The idea underlying correlated pseudo-marginal methods (CPM), as initially proposed in \cite{dahlin2015accelerating} and \cite{deligiannidis2015correlated}, is that having correlated likelihoods will reduce the stochastic variability in the acceptance ratio. This reduces the stickiness in the MCMC chain, which is typically due to excessively varying likelihood approximations, when these are obtained using a ``too small'' number of Monte Carlo draws (named ``particles''). In fact, while the variability of these estimates can be mitigated by increasing the number of particles, of course this has negative consequences on the computational budget. Instead CPM strategies allow for considerably smaller number of particles when trying to alleviate the stickiness problem, see for example \cite{golightly2018correlated} for applications to stochastic kinetic models, and \cite{wiqvist2021efficient} and \cite{botha2021particle} for stochastic differential equation mixed-effects models. 
Interestingly, implementing CPM approaches is trivial. \cite{deligiannidis2015correlated} and \cite{dahlin2015accelerating} correlate the estimated likelihoods at the proposed and
current values of the model parameters by correlating the underlying standard normal
random numbers used to construct the estimates of the likelihood, via a Crank-Nicolson proposal. We found particular benefit with the ``blocked'' PM approach (BPM) of \cite{tran2016block} (see also \citealp{choppala2016bayesian} for inference in state-space models), which we now describe in full generality, i.e. regardless of our synthetic likelihoods approach which is instead considered later. 

Denote with $\mathrm{U}$ the vector of all ``auxiliary variables'', i.e. pseudorandom numbers (typically standard Gaussian or uniform) that are
necessary to produce a non-negative unbiased likelihood approximation $\hat{p}(y|\theta,\mathrm{U})$ at a given parameter $\theta$ for data $y$. Notice $U$ should contain the pseudo-random numbers that are used to ``forward simulate'' from a model, but can include also other auxiliary variables, for example the pseudo-random numbers that are generated when performing the resampling step in sequential Monte Carlo. In \cite{tran2016block} the set $\mathrm{U}$  is divided into $G$ blocks $U=(U_{(1)},...,U_{(G)})$, and one of these blocks is updated jointly with $\theta$ in each MCMC iteration as described below. Let $\hat{p}(y|{\theta},\mathrm{U}_{(i)})$ be the estimated unbiased likelihood obtained using the $i$th block of random variates $U_{(i)}$, $i=1,...,G$. Define the joint posterior of $\theta$ and $\mathrm{U}$  as
\begin{align}
\pi({\theta},\mathrm{U}|y) & \propto {\hat{p}}(y|\theta,\mathrm{U})
\pi(\theta)\prod_{i=1}^G
p_{U}(\mathrm{U}_{(i)})\label{eq:augmented-posterior}\\
\intertext{where $\theta$ and $\mathrm{U}$ are a-priori independent and}
{\hat{p}}(y|\theta,\mathrm{U}) &:=\frac1G\sum_{i=1}^G\hat{p}(y|{\theta},
\mathrm{U}_{(i)})\label{eq:unbiased-average}
\end{align}
is the average of the $G$ unbiased likelihood estimates and hence also unbiased.
We then update the parameters jointly with a randomly-selected block $\mathrm{U}_{\left(K\right)}$ in each MCMC iteration,
with $\Pr\left(K=k\right)=1/G$ for any $k=1,...,G$. ``Updating a randomly selected block'' means that only for that picked block $U_{(k)}$ new pseudorandom values are produced (and hence are ``refreshed'') while for the other blocks these variates are kept fixed to the previously accepted values.
Using this scheme, the acceptance probability for a joint move from the current set $(\theta^c,\mathrm{U}^c)$ to a proposed set $(\theta^p,\mathrm{U}^p)$ generated using some proposal function $g(\theta^p,\mathrm{U}^p|\theta^c,\mathrm{U}^c)=g(\theta^p|\theta^c)g(\mathrm{U}^p|\mathrm{U}^c)$, is
\begin{equation}
\alpha = \min\left\{ 1,\frac{{\hat{p}}\left(y|{\theta}^{p},\mathrm{U}_{\left(1\right)}^{c},...,\mathrm{U}_{\left(k-1\right)}^{c},\mathrm{U}_{\left(k\right)}^{p},\mathrm{U}_{\left(k+1\right)}^{c},...,\mathrm{U}_{\left(G\right)}^{c}\right)\pi\left({\theta}^{p}\right)}{{\hat{p}}\left(y|{\theta}^{c},\mathrm{U}_{\left(1\right)}^{c},...,\mathrm{U}_{\left(k-1\right)}^{c},\mathrm{U}_{\left(k\right)}^{c},\mathrm{U}_{\left(k+1\right)}^{c},...,\mathrm{U}_{\left(G\right)}^{c}\right)\pi\left({\theta}^{c}\right)}\frac{g\left({\theta}^{c}|{\theta}^{p}\right)}{g\left({\theta}^{p}|{\theta}^{c}\right)}\right\}.
\label{eq:acceptance prob}
\end{equation}
Hence in case of proposal acceptance we update the joint vector $(\theta^c,\mathrm{U}^c):=(\theta^p,\mathrm{U}^p)$ and move to the next iteration, where $\mathrm{U}^p=(\mathrm{U}_{\left(1\right)}^{c},...,\mathrm{U}_{\left(k-1\right)}^{c},\mathrm{U}_{\left(k\right)}^{p},\mathrm{U}_{\left(k+1\right)}^{c},...,\mathrm{U}_{\left(G\right)}^{c})$.
The resulting chain targets \eqref{eq:augmented-posterior} \citep{tran2016block}.
Notice the random variates used to compute the likelihood at the numerator of \eqref{eq:acceptance prob} are the same ones as for the likelihood at the denominator except for the $k$-th block, hence $G-1$ blocks are shared between the numerator and denominator. Perturbing only a small fraction of the pseudo-random numbers  induces beneficial correlation between subsequent pairs of likelihood estimates, as in this case the variance of $\alpha$ gets smaller compared to having all entries in $U$ getting updated at each iteration. Also, we considered $g(\mathrm{U}^p|\mathrm{U}^c)\equiv p_U(U_{(k)}^p)$ hence the simplified expression \eqref{eq:acceptance prob}.
The correlation between $\log\hat{p}\left(y|{\theta}^{p},\mathrm{U}^{p}\right)$ and $\log\hat{p}\left(y|{\theta}^{c},\mathrm{U}^{c}\right)$ is approximately $\rho=1-1/G$ \citep{tran2016block}, so the larger the number of groups $G$ that can be formed and the higher the correlation (at least theoretically). Also, note that the $G$ approximations $\hat{p}(y|{\theta},
\mathrm{U}_{(i)})$ can be run in parallel on multiple processors when these likelihoods are approximated using particle filters. 

We now consider synthetic likelihoods. Denote with $\mathrm{U}_j$ the vector of auxiliary variables employed when producing the $j$-th model simulation ($j=1,...,M$). Denote with $(U_1,...,U_M)$ the vector stacking the variates generated across all $M$ model simulations. We distribute those variates across $G$ blocks: assume for simplicity that $M$ is a multiple of $G$, so that for example the $i$-th block $U_{(i)}$  could be the collection of the pseudo-random numbers used in a small subset of the $M$ model simulations, so that $\sum_{i=1}^G \dim(U_{(i)})=\dim(U_1,...,U_M)$ and $U_{(i)}\bigcap U_{(i')}=\{\emptyset\}$, $i\neq i'$. That is $(U_{(1)},...,U_{(G)})$ is a partition of $(U_1,...,U_M)$.

Same as before, in each MCMC iteration we ``refresh'' the variates from a randomly sampled block, while the other variates are kept fixed to the previously accepted values. 
In our synthetic likelihood approach we do not make use of \eqref{eq:unbiased-average} and take instead $p(s|\theta,\mathrm{U})$ without decomposing this into a sum of $G$ contributions. We do not in fact compute separately the $p(s|\theta,\mathrm{U}_{(i)})$, since we found that in order for each $p(s|\theta,\mathrm{U}_{(i)})$ to behave in a numerically stable way, a not too small number of simulations $M_{(i)}$ should be devoted for each sub-likelihood term, or otherwise the corresponding estimated covariance may misbehave (e.g., may result not positive-definite). Therefore, in practice, we just obtain the joint $p(s|\theta,\mathrm{U})$, and \eqref{eq:acceptance prob} becomes
\begin{equation}
\alpha = \min\left\{ 1,\frac{{{p}}\left(s|{\theta}^{p},\mathrm{U}_{\left(1\right)}^{c},...,\mathrm{U}_{\left(k-1\right)}^{c},\mathrm{U}_{\left(k\right)}^{p},\mathrm{U}_{\left(k+1\right)}^{c},...,\mathrm{U}_{\left(G\right)}^{c}\right)\pi\left({\theta}^{p}\right)}{{{p}}\left(s|{\theta}^{c},\mathrm{U}_{\left(1\right)}^{c},...,\mathrm{U}_{\left(k-1\right)}^{c},\mathrm{U}_{\left(k\right)}^{c},\mathrm{U}_{\left(k+1\right)}^{c},...,\mathrm{U}_{\left(G\right)}^{c}\right)\pi\left({\theta}^{c}\right)}\frac{g\left({\theta}^{c}|{\theta}^{p}\right)}{g\left({\theta}^{p}|{\theta}^{c}\right)}\right\},
\label{eq:corrsyn-acceptance prob}
\end{equation}
which we therefore call ``correlated synthetic likelihood'' (CSL) approach.
From the analytic point of view our correlated likelihood $p(s|\theta,\mathrm{U})$ is the same unbiased approximation given in \cite{price2018bayesian} (also in Supplementary Material), hence CSL uses the BSL approach, the only difference with BSL being that the numerator and denominator of \eqref{eq:corrsyn-acceptance prob} have $G-1$ blocks in common, while in BSL all pseudo-random numbers are refreshed at each iteration for each new likelihood. 

In our experiments we show that using the acceptance criterion \eqref{eq:corrsyn-acceptance prob} into Algorithm \ref{alg:synlikMCMC} (regardless of the use of our ASL proposal kernel) is of benefit to ease convergence and also increase chains mixing. Moreover, it comes with no computational overhead compared to not using correlated synthetic likelihoods. The only potential issue would be some careful extra coding and the need to store $(\mathrm{U}_{(1)},...,\mathrm{U}_{(G)})$ in memory, which could be large dimensional with complex model simulators.

\section{Algorithmic initialization using BOLFI and ELFI}\label{sec:bolfi}

This section does not contain novel material, but it is useful to inform modellers using SL approaches of alternative strategies to initialize SL algorithms.
We consider the case where obtaining a reasonable starting value $\theta_1$ for $\theta$ by trial-and-error is not feasible, due to the computational cost of evaluating the SL density at many candidates for $\theta_1$. At minimum, we need to find a value $\theta_1$ such that the corresponding SL density (the biased $p_M$ or the unbiased one in the sense of \citealp{price2018bayesian}) has a positive definite covariance matrix $\hat{\Sigma}$. This is not ensured when the summaries are simulated from highly non-representative values of $\theta$, which would result in an MCMC algorithm that halts. The issue is critical, as testing many values $\theta_1$ can be prohibitively expensive, both because the dimension of $\theta$ can be large and because the model itself might be slow to simulate from. 

An approach developed in \cite{gutmann2016bayesian} uses Bayesian optimization  when the likelihood function is intractable but realizations from a stochastic model simulator are available, which is exactly the framework that applies to ABC and SL. The resulting method, named BOLFI (Bayesian optimization for likelihood-free inference), locates a $\theta$ that either minimizes the expected value of $\log \Delta$, where $\Delta$ is some discrepancy between simulated and observed summary statistics, say $\Delta=\parallel s^* - s \parallel$ for some distance $\parallel \cdot \parallel$, or alternatively can be used to minimize the negative log-SL expression. For example, $\parallel \cdot \parallel$ could be the Euclidean distance $((s^* - s)'(s^* - s)')^{1/2}$, or a Mahalanobis distance $((s^* - s)'A(s^* - s)')^{1/2}$ for some square matrix $A$ weighting the individual contributions of the entries in $s^*$ and $s$ (see \citealp{prangle2017adapting}). The appeal of BOLFI is that (i) it is able to rapidly focus the exploration in those regions of the parameter space where either $\Delta$ is smaller, or the SL is larger, and (ii) it is implemented in \texttt{ELFI} \citep{lintusaari2018elfi}, the Python-based open-source engine for likelihood-free inference.

Hence, when dealing with expensive simulators, BOLFI is ideally positioned to minimize the number of attempts needed to obtain a reasonable value $\theta_1$, to be used to initialize the synthetic likelihoods approach. BOLFI replaces the expensive realizations from the model simulator with a ``surrogate simulator'' defined by a Gaussian process (GP, \citealp{rasmussen2004gaussian}). Using simulations from the actual (expensive) simulator to form a collection of pairs such as $(\theta,\log\Delta)$, the GP is trained on the generated pairs and the actual optimization in BOLFI only uses the computationally cheap GP simulator. This means that the optimum returned by BOLFI does not necessarily reflect the best $\theta$ generating the observed $s$. It is possible to use the BOLFI optimum to initialize some other procedure within \texttt{ELFI}, such as Hamiltonian Monte Carlo via the NUTS algorithm of \cite{hoffman2014no}. However, the \texttt{ELFI} version of NUTS uses, again, the GP surrogate of the likelihood function. Once the BOLFI optimum is obtained, it can be used to initialise (B)SL MCMC which still uses simulations from the true model, and these may be expensive, but at least are initialised at a $\theta$ which should be ``good enough''  to avoid a long and expensive burnin. Illustrations of BOLFI are in sections \ref{sec:gk_bolfi} and \ref{sec:astro-simul}. A more recent contribution, exploiting GPs to predict a log-SL, is in \cite{jarvenpaa2020parallel}.

\section{Simulation studies}\label{sec:simulation-studies}

Here follow four simulation studies. A fifth one, using a ``perturbed'' $\alpha$-stable model built to pose a challenge to CSL, is in Supplementary Material.
In all the considered examples we use $N=1$, i.e. the ASL proposal kernel is updated at each iteration. Within ASL we always use a Gaussian proposal based on \eqref{eq:mu-conditional-general}--\eqref{eq:cov-conditional-general}, and never the multivariate Student's one.

\subsection{g-and-k distribution}\label{sec:g-and-k}

The g-and-k distribution is a standard toy model for case studies having intractable likelihoods (e.g. \citealp{allingham2009bayesian,fearnhead2012constructing}), in that its simulation is straightforward, but it does not have a closed-form probability density function (pdf). Therefore the likelihood is analytically intractable. However, it has been noted in \cite{rayner2002numerical} that one can still numerically compute the pdf, by 1) numerically inverting the quantile function to get the cumulative distribution function (cdf), and 2)  numerically differentiating the cdf, using finite differences, for instance. Therefore ``exact'' Bayesian inference (exact up to numerical discretization) is possible. This approach is implemented in the \texttt{gk} R package \citep{gk}.

The  $g$-and-$k$ distributions is a flexibly shaped distribution that is used to model non-standard data
through a small number of parameters. It is defined by its quantile
function, see \cite{gk} for an overview. Essentially, it is possible to generate a draw $Q$ from the distribution using the following scheme
\begin{equation}
Q= A+B\biggl[1+c\frac{1-\exp(-g\cdot u)}{1+\exp(-g\cdot u)}\biggr](1+u^2)^k\cdot u
\label{eq:g-k-inverse}
\end{equation}
where $u\sim N(0,1)$, $A$ and $B$ are location and scale parameters and $g$ and $k$ are related to skewness and kurtosis. Parameters restrictions are $B>0$ and $k>-0.5$. 
We assume $\theta=(A,B,g,k)$ as parameter of interest, by noting that it is customary to keep $c$ fixed to $c=0.8$ (\citealp{drovandi2011likelihood,rayner2002numerical}). We use the summaries $s(w)=(s_{A,w},s_{B,w},s_{g,w},s_{k,w})$ suggested in \cite{drovandi2011likelihood}, where $w$ can be  observed and simulated data $y$ and $y^*$ respectively:
\begin{align*}
s_{A,w}&=P_{50,w} & s_{B,w}&=P_{75,w}-P_{25,w},\\ 
s_{g,w}&=(P_{75,w}+P_{25,w}-2s_{A,w})/s_{B,w} & s_{k,w}&=(P_{87.5,w}-P_{62.5,w}+P_{37.5,w}-P_{12.5,w})/s_{B,w}
\end{align*}
where $P_{q,w}$ is the $q$th empirical percentile of $w$. That is $s_{A,w}$ and $s_{B,w}$ are the median and the inter-quartile range of $w$ respectively.
We use the simulation strategy outlined above to generate data $y$, consisting of $1,000$ independent samples from the g-and-k distribution using parameters $\theta=(A,B,g,k)=(3,1,2,0.5)$. We place uniform priors on the parameters: $ A\sim U(-30,30)$, $ B\sim U(0,30)$, $g\sim U(0,30)$, $k\sim U(0,30)$. 

\begin{figure}[ht]
\centering
\begin{subfigure}[b]{0.49\textwidth}
\includegraphics[width=\textwidth]{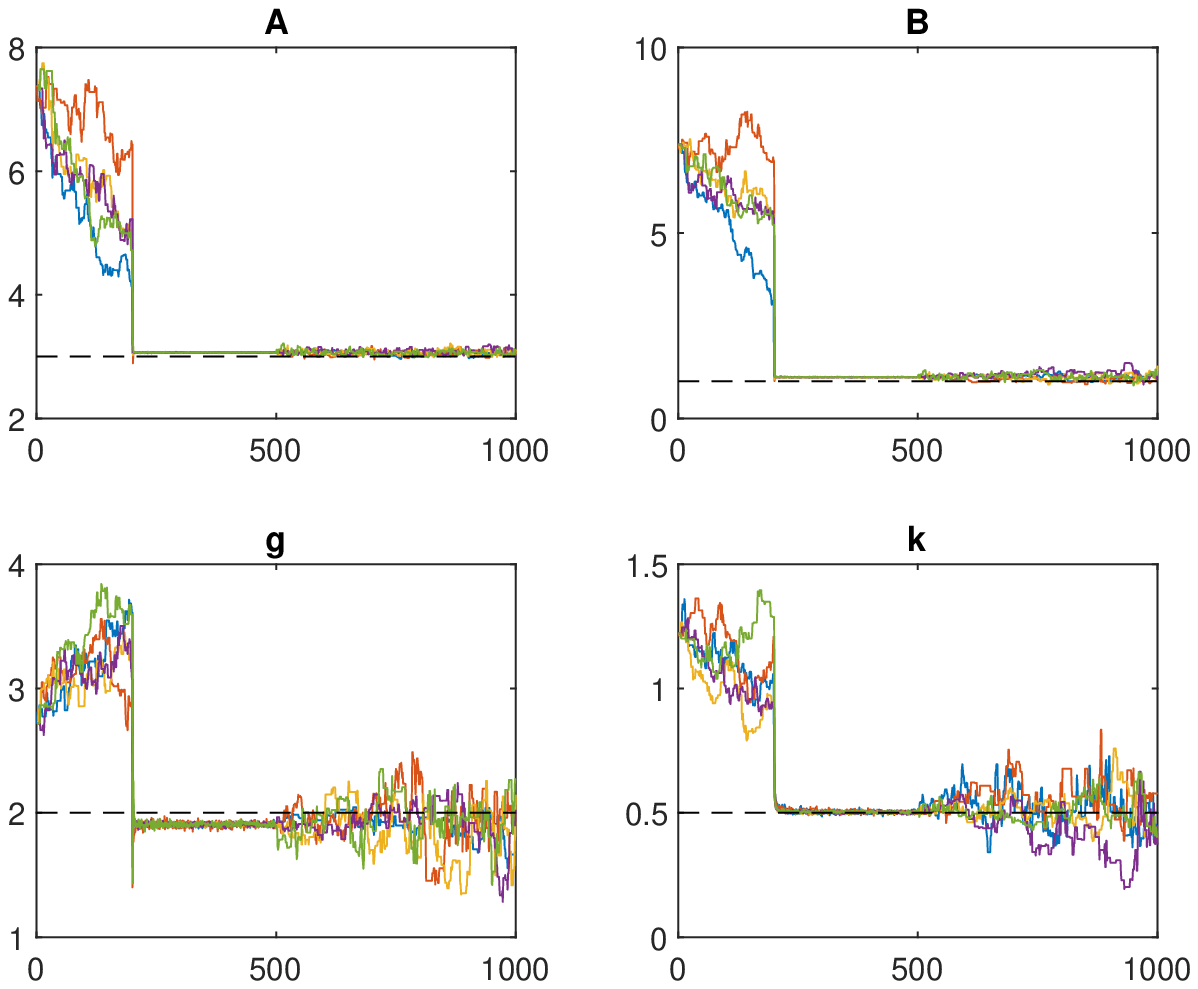}
\caption{}\label{fig:g-and-k_alltraceplots}
\end{subfigure}
\begin{subfigure}[b]{0.49\textwidth}
\includegraphics[width=\textwidth]{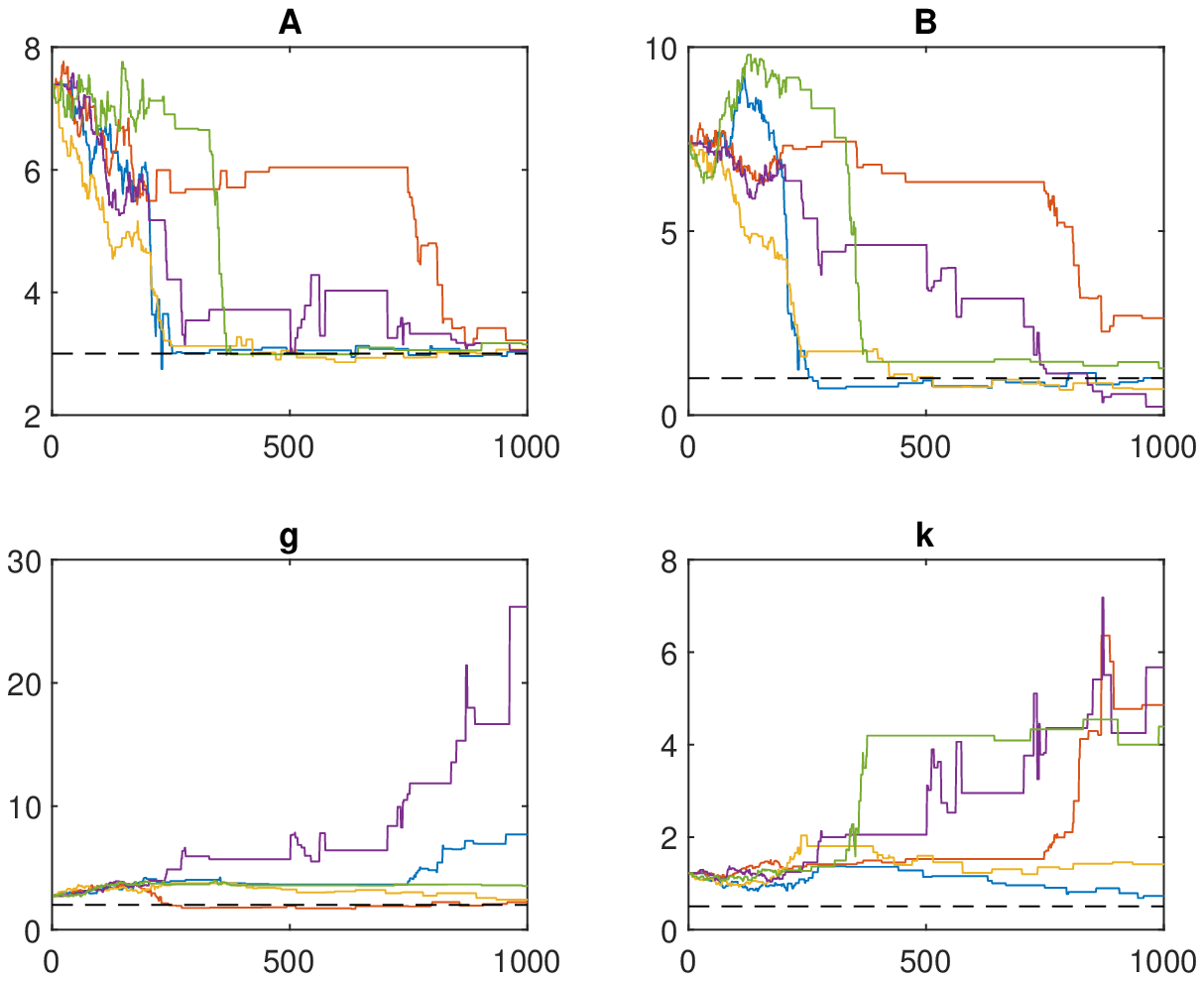}
\caption{}\label{fig:g-and-k_Haario}
\end{subfigure}
\caption{g-and-k: (a)  five inference runs using BSL with ASL sampler employed from iteration 200 to 500. We display the first 1,000 iterations to emphasize the effect of the ASL adaption. The black dashed lines mark ground-truth parameters. (b) five inference runs of BSL (only the first 1,000 iterations are displayed for comparison with Figure \ref{fig:g-and-k_alltraceplots}). Here BSL uses the sampler of \cite{haario2001adaptive} from iteration 200 onward.}
\end{figure}

We run five inference attempts independently, always starting at $\theta_0=(7.389,    7.389,2.718,1.221)$ and using the same data. For all experiments,  $M=1,000$ model simulations are produced at each proposed parameter and we found this value of $M$ to be necessary given the used starting parameters, or we would not collect enough parameter moves.
However if we instead initialize the simulations close to the true values then a considerably smaller $M$ can be employed (this is discussed later), which shows that our contribution on accelerating convergence to the high posterior probability region is important.
We start by running  $K=200$ burnin iterations, during which we advance the chain by proposing parameters using a Gaussian random walk acting on log-scale, i.e. on $\log\theta$, with a constant diagonal covariance matrix having standard deviations (on log-scale) given by $[0.025, 0.025, 0.025, 0.025]$ for $(\log A,\log B,\log g,\log k)$ respectively. Given the short burnin, in the first $K$ iterations we implement  a Markov-chain-within-Metropolis strategy (MCWM, \citealp{andrieu2009pseudo}) to increase the mixing of the algorithm before our sequentially guided ASL strategy starts (shortly, MCWM differs from a standard Metropolis-Hastings algorithm in that the stochastic likelihood approximation in the denominator of the Metropolis-Hastings ratio is re-evaluated at the last accepted parameter value, instead of using the value of the previously accepted synthetic likelihood). Notice the use of MCWM is strictly limited to the burnin iterations, since MCWM doubles the execution time and its theoretical properties are not well understood. At iteration $K+1$, our ASL Algorithm \ref{alg:adaptiveSL} starts and is let run for 300 iterations (notice a much smaller number of iterations than 300 can be used, say 50. We chose 300 for pictorial reasons as the effect of ASL gets better noticed in figures).   Afterwards BSL inherits the last draw accepted by ASL and reverts to using the adaptive Metropolis random walk proposal of \cite{haario2001adaptive}, thereafter denoted ``Haario'', which is used for further 2,800 iterations and is adapted as described in Supplementary Material. Therefore the total length of the chain is 3,300 ($K=200$ iterations, then 300 ASL iterations then 2,800 further iterations). The covariance matrix in the adaptive proposal of ``Haario'' is updated every 30 iterations.  The five independent inference attempts are in Figure \ref{fig:g-and-k_alltraceplots}. We notice that during the burnin the chains are still quite far from the ground truth. However, as soon as ASL kicks in (iteration 201), we notice a large jump towards the true parameters. The proposal in the ASL algorithm produces a high acceptance rate which although it induces very local moves, it never gets stuck and thus provides useful information to initialize the covariance matrix in ``Haario''. 

We now avoid using our guided ASL and run BSL using again MCWM during the burnin iterations and the adaptive ``Haario'' strategy for the remaining iterations, with results in Figure \ref{fig:g-and-k_Haario}. This shows the difficulty of running BSL when starting parameters are in the tails of the posterior, where several runs completely fail and only occasionally they manage to reach the bulk of the posterior. Furthermore, the patterns are very sticky. This is because the adaptive ``Haario'' proposal tunes the covariance of the Gaussian random walk sampler on the previous history of the chain, which becomes problematic if many rejections occur, as in this case the covariance shrinks, thus making the recovery difficult. This is why the high acceptance rate of ASL, coupled to the rapid convergence towards the posterior's bulk, helps collecting moves that are useful for the learning of the covariance matrix for the adaptive random walk proposal.

We mentioned that if a starting parameter value is chosen closer to the ground truth value a smaller $M$ can be employed (and recall from Figure \ref{fig:g-and-k_Haario} that with a standard adaptive proposal method BSL failed even with $M=1,000$). As an example, we take the sample mean of the acceptances produced via ASL from iteration 200 to 500, and use this sample mean to initialize BSL when using the ``Haario'' proposal with as little as $M=50$: the mixing of BSL in this case is very satisfactory (results shown in the Supplementary Material) and actually even using $M=10$ allows good mixing but slightly worse tail performance. Therefore, ASL can be really useful in enabling standard BSL to be used with considerable computational savings (5,000 iterations of BSL can be performed in under a minute when $M=50$).

\subsubsection{Using correlated synthetic likelihood without ASL}

\begin{figure}[ht]
    \centering
    \includegraphics[scale=0.5]{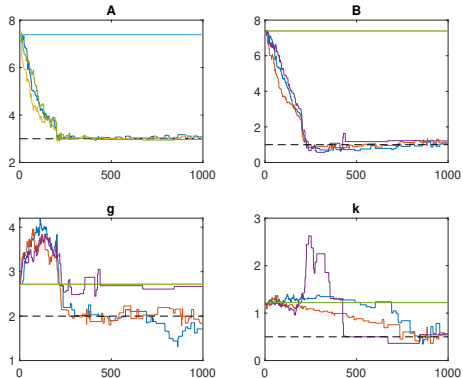}
    \caption{g-and-k: 1,000 iterations from CSL, using  $G=100$ groups. The black dashed lines mark ground-truth parameters. Solid horizontal lines correspond to failed attempts.}
    \label{g-k_CSL_numgroups=100}
\end{figure}

Here we consider the correlated synthetic likelihood (CSL) approach outlined in section \ref{sec:corrsynlik}, without the use of our ASL approach for proposing parameters, to better appreciate the individual effect of using correlated likelihoods. Notice \eqref{eq:g-k-inverse} immediately suggests how to implement CSL, since the $u$ appearing in \eqref{eq:g-k-inverse} can be thought as a scalar realization of the $U$ variate in section \ref{sec:corrsynlik}.
We initialised parameters at the same starting values as in the previous experiments, across five independent inference attempts.
We used CSL throughout, including the burnin phase, that is we do not employ MCWM during the burnin.   
After 200 burnin iterations with fixed covariance matrix, we propose parameters using ``Haario''. 
We illustrate results obtained with $G=100$ blocks, which should imply a theoretical correlation of $\rho=1-1/100=0.99$ between estimated synthetic loglikelihoods, see Figure \ref{g-k_CSL_numgroups=100}. Figure \ref{g-k_CSL_numgroups=100} shows that, while two attempts failed (essentially because the 200 burnin iterations did not produce any acceptance), the remaining attempts managed to reach the ground-truth values. Recall that when using BSL without induced correlation (and employing the same ``Haario'' proposal sampler) we produced Figure \ref{fig:g-and-k_Haario}. The benefits of recycling pseudo-random variates are noticeable. Similar plots, but using $G=50$, are in Supplementary Material. The comparison between the two cases $G=50$ and $G=100$ shows that inducing higher correlation (i.e. $G=100$) allow faster convergence to ground-truth parameters, however at the same time the mixing is reduced due to a reuse of perhaps too many $U$-variates, whereas with $G=50$ the chains appear to mix better.

\subsubsection{Initialization using ELFI and BOLFI}\label{sec:gk_bolfi}

Here we show results from the BOLFI optimizer (discussed in section \ref{sec:bolfi}) to find a promising area in the posterior region and hence provide a useful starting value for SL. We use the \texttt{ELFI} software. In this particular example BOLFI uses a Gaussian Process (GP) to learn the possibly complex and nonlinear relationship between discrepancies (or log-discrepancies) $\log\Delta$ and corresponding parameters $\theta$. 
We found that for this specific example, where we set very wide and vague priors, we could not infer the parameters using BOLFI with the LCB (lower confidence bound) acquisition function regardless the value set for $J_1$. This is because while in previous inference attempts we used MCMC methods to explore the posterior and having very vague priors was still feasible, here having initial samples provided by very uninformative priors is not manageable. In this section we use $A\sim U(-10,10)$, $B\sim U(0,10)$, $g\sim U(0,10)$, $k\sim U(0,10)$. These priors are narrower than in previous attempts but are still wide and uninformative enough to make this experiment interesting and challenging.

\begin{figure}[ht]
    \centering
    \begin{subfigure}[b]{0.7\textwidth}
        \includegraphics[width=\textwidth]{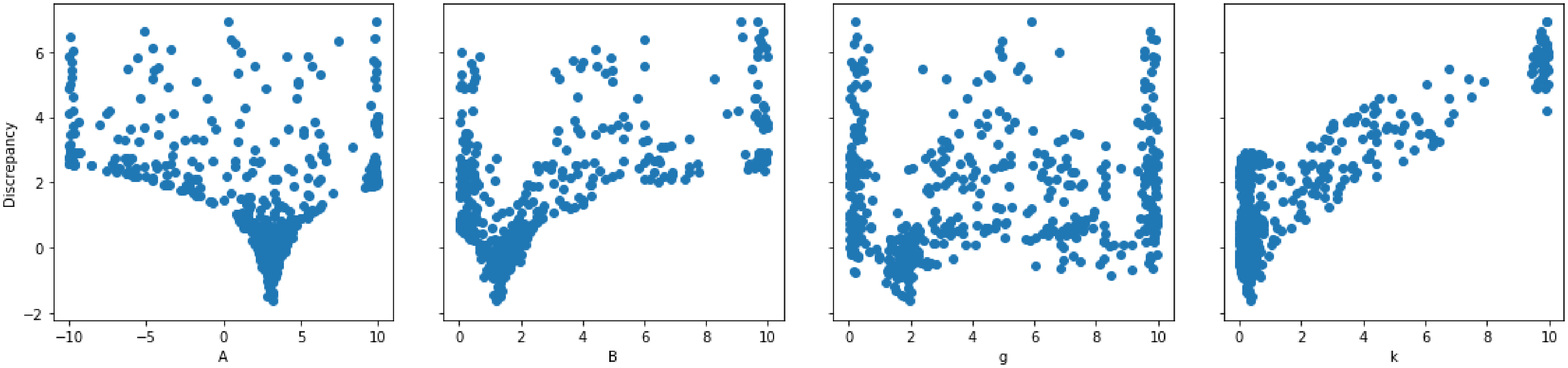}
        \caption{}
    \end{subfigure}
    
    \begin{subfigure}[b]{0.7\textwidth}
        \includegraphics[width=\textwidth]{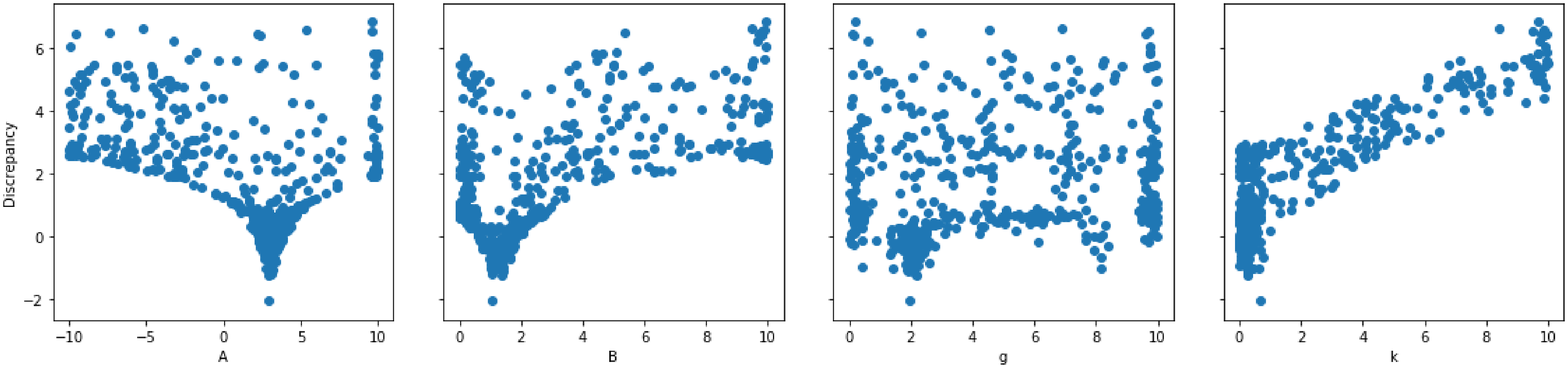}
        \caption{}
    \end{subfigure}
    \caption{$g$-and-$k$: log-discrepancies for the tested parameters using BOLFI with $J_1=20$ (top) and $J_1=100$ (bottom). From left to right: plots for $A$, $B$, $g$ and $k$ respectively.}
    \label{fig:g-and-k_bolfi}
\end{figure}

Once the $J_1$ training samples are obtained, BOLFI starts optimizing parameters by iteratively fitting a GP and proposing points $\theta_{(j)}$ such that each $\theta_{(j)}$ attempts at reducing $\log\Delta$, $j=1,...,J_2$. We first consider $J_2=500$ and then $J_2=800$.  The clouds of points in Figure \ref{fig:g-and-k_bolfi}  represent all $J_1+J_2$ values of log-discrepancies $\log\Delta$ (for $(J_1,J_2)=(20,500)$ and $(J_1,J_2)=(100,500)$) and corresponding parameter values. The smallest values of $\log\Delta$ cluster around the ground-truth parameters which we recall are $A=3$, $B=1$, $g=2$, $k=0.5$. The values of the optimized discrepancies are in Supplementary Material. Even with a very small $J_1$ the obtained results appear very promising. Also, even though the estimates for $k$ seem to be bounded by the lower limit we set for its prior, we can clearly notice a trend, in that smaller values for $k$ return smaller discrepancies. BOLFI can be an effective tool to initialize an MCMC procedure for synthetic likelihoods.

\subsection{Supernova cosmological parameters estimation with twenty summary statistics}

We present an astronomical example taken from \cite{jennings2017astroabc}. There, the ABC algorithm by \cite{BeaumontEtAl2009} was used for likelihood-free inference. The algorithm in \cite{BeaumontEtAl2009} is a sequential Monte Carlo (SMC) sampler, hereafter denoted ABC-SMC, which propagates many parameter values (``particles'') through a sequence of approximations of the posterior distribution of the parameters. The sequence of approximations depends on the sequence of tolerances  $\epsilon_{1:T}$, where $T$ is the final iteration of the procedure. The different approaches used in order to create the series of decreasing tolerances \citep{BeaumontEtAl2009, del2012adaptive}, together with the choice for $T$, can lead to inefficient sampling \citep{simola2020adaptive}. For this reason, rather than using the ABC-SMC algorithm, we employed one of its extensions, the Adaptive ABC--PMC (hereafter aABC--PMC) found in \cite{simola2020adaptive}. When using the aABC--PMC algorithm both the series of decreasing tolerances and $T$ are automatically selected, by looking at the online behaviour of the approximations to the posterior distribution (aABC--PMC is  also implemented in ELFI). Our goal is to show how synthetic likelihoods may be as well used in order to tackle the inferential problem, and a comparison with aABC--PMC and BOLFI is presented. In \cite{jennings2017astroabc} the analysis relied on the SNANA light curve analysis package \citep{kessler2009snana} and its corresponding implementation of the SALT--II light curve fitter presented in \cite{guy2010supernova}. A sample of $400$ supernovae with redshift range $z \in [0.5, 1.0]$ are simulated and then binned into $20$ redshift bins.  However, for this example, we did not use SNANA and data is instead simulated  following the procedure in Supplementary Material. The model that describes the distance modulus as a function of redshift $z$, known in the astronomical literature as Friedmann--Robertson--Model \citep{condon2018lambdacdm}, is:  

\begin{equation}
\mu_{i}(z_{i};\Omega_{m},\Omega_{\Lambda},\Omega_{k},w_{0}, h_0) \propto 5\log_{10}\biggl(\frac{c(1+z_i)}{h_{0}}\biggr)\int_{0}^{z_i}\frac{dz'}{E(z')},
\label{eq:supernovaFM}
\end{equation}
where $E(z) = \sqrt{\Omega_{m} (1+z)^{3} + \Omega_{k} (1+z)^{2} + \Omega_{\Lambda}e^{3\int_{0}^{z} dln(1+z')[1+w(z')]}}$.

The cosmological parameters involved in \eqref{eq:supernovaFM} are five. The first three parameters are the matter density of the universe, $\Omega_m$, the dark energy density of the universe, $\Omega_{\Lambda}$ and the radiation and relic neutrinos, $\Omega_{k}$. A constraint is involved when dealing with these three parameters, which is $\Omega_m+\Omega_{\Lambda}+\Omega_{k}=1$ \citep{genovese2009inference, tripathi2017dark, usmani2008dark}. The final two parameters are, respectively, the present value of the dark energy equation, $w_0$, and the Hubble constant, $h_0$. A common assumption involves a flat universe, leading to $\Omega_{k}=0$, as shown in \cite{tripathi2017dark,usmani2008dark}. As a result, \eqref{eq:supernovaFM} simplifies and in particular $E(z)$ can be written as $E(z) = \sqrt{\Omega_{m} (1+z)^{3} + (1-\Omega_{m})e^{3\int_{0}^{z} dln(1+z')[1+w(z')]}}$, where we note that $\Omega_{\Lambda}= 1- \Omega_{m}$. Same as in \cite{jennings2017astroabc}, we work under the flat universe assumption.
Concerning the Dark Energy Equation of State (EoS), $w(\cdot)$, we use the parametrization proposed in \cite{chevallier2001accelerating} and in \cite{linder2003exploring}:
\begin{equation}
w(z)=w_0 + w_{a} (1 - a) = w_0 + w_{a} \frac{z}{1+z}.
\label{eq:supernova.eos}
\end{equation}
According to \eqref{eq:supernova.eos}, $w$ is assumed linear in the scale parameter. Another common assumption relies on $w$ being constant; in this case $w=w_0$. We note that several parametrizations have been proposed for the EoS (see for example \cite{huterer2001probing}, \cite{wetterich2004phenomenological} and \cite{usmani2008dark}). For the present example, ground-truth parameters are set as follows: $\Omega_m = 0.3$, $\Omega_k = 0$, $w_0 = -1.0$ and $h_0 = 0.7$.

In the present study $h_0$ is assumed known. Similarly to \cite{jennings2017astroabc},  we aim at inferring the cosmological parameters $\theta=(\Omega_m,w_0)$. The distance function used to compare $\mu$ with the ``simulated'' data $\mu_{sim}(z)$ is:
\begin{equation}
\rho(\mu,\mu_{sim}(z)) = \sum_{i} {(\mu_{i} - \mu_{sim}(z_{i}))^{2}}.
\label{eq:supernova_distance}
\end{equation}

 We recall that the aABC--PMC algorithm in \cite{simola2020adaptive} uses a series of automatically selected decreasing tolerances $\epsilon_{1:T}$, each inducing a better approximation to the true posterior distribution as $t\in[1,T]$ increases. When the stopping rule, based on the improvement between two consecutive posterior distributions is satisfied, the procedure automatically halts. While the ABC posterior based on $\epsilon_1$ uses the prior distribution as proposal function, for $t>1$ the aABC--PMC uses the previous iteration's ABC posterior to produce candidates, just like regular ABC-PMC or other sequential ABC procedures. In this work, as done also by \cite{jennings2017astroabc}, we follow  \cite{BeaumontEtAl2009} regarding the selection of the perturbation kernel, which is a Gaussian distribution centered to the selected particle and having variance equal to twice the weighted sample variance of the particles selected in the previous iteration. 
The specifications for the aABC--PMC algorithm are found in \cite{simola2020adaptive}.
For all experiments, we set priors $\Omega_m \sim Beta(3, 3)$, since $\Omega_m$ must be in $(0,1)$, and $w_0 \sim \mathcal{N}(-0.5, 0.5^2)$.

 \subsubsection{Inference} \label{sec:astro-simul}
 
We describe how to forward simulate from the model in Supplementary Material. 
We take $s=(\mu_1,...,\mu_{20})$ as ``observed'' summary statistics corresponding to the stochastic input generated as described in Supplementary Material. Notice, in our case $s$ is the trivial summary statistic, in that $(\mu_1,...,\mu_{20})$ is the data itself.
We investigate the assumption in the Supplementary Material and find that this is statistically supported, at least for summaries simulated at ground-truth parameter values. However, notice that a different behaviour might occur at other values of $\theta$, for example at those values far from the ground truth. We found it impractical to consider $M$ in the order of thousands, however using a smaller value of, say, $M=100$ would produce an ill-conditioned covariance matrix. To overcome this problem we found it essential to use a shrinkage estimator of  $\hat{\Sigma}_{M,\theta}$, such as the one due to \cite{warton2008penalized} and employed in a BSL context in \cite{nott2019bayesian}. This way we managed to use as little as $M=100$ model simulations.
In this section we denote the BSL approach using shrinkage as ``sBSL''. We compare sBSL with the correlated synthetic likelihoods approach plugged into ASL, and denote this method ``ACSL'' (we employed shrinkage also within ACSL). We always use $M=100$, and within ACSL we experiment with several number of blocks, namely $G=5$ and 10. For all methods, starting parameter values are $(\Omega_m=0.90, w_0= -0.5)$, see the Supplementary Material for further details on the MCMC settings. We first note that sBSL is unable to move away from the starting parameter values, and hence this attempt is a failure. Introducing correlation between synthetic loglikelihoods is a key feature for the success of ACSL in this case study.

\begin{figure}[ht]
   \centering
\includegraphics[scale=0.4]{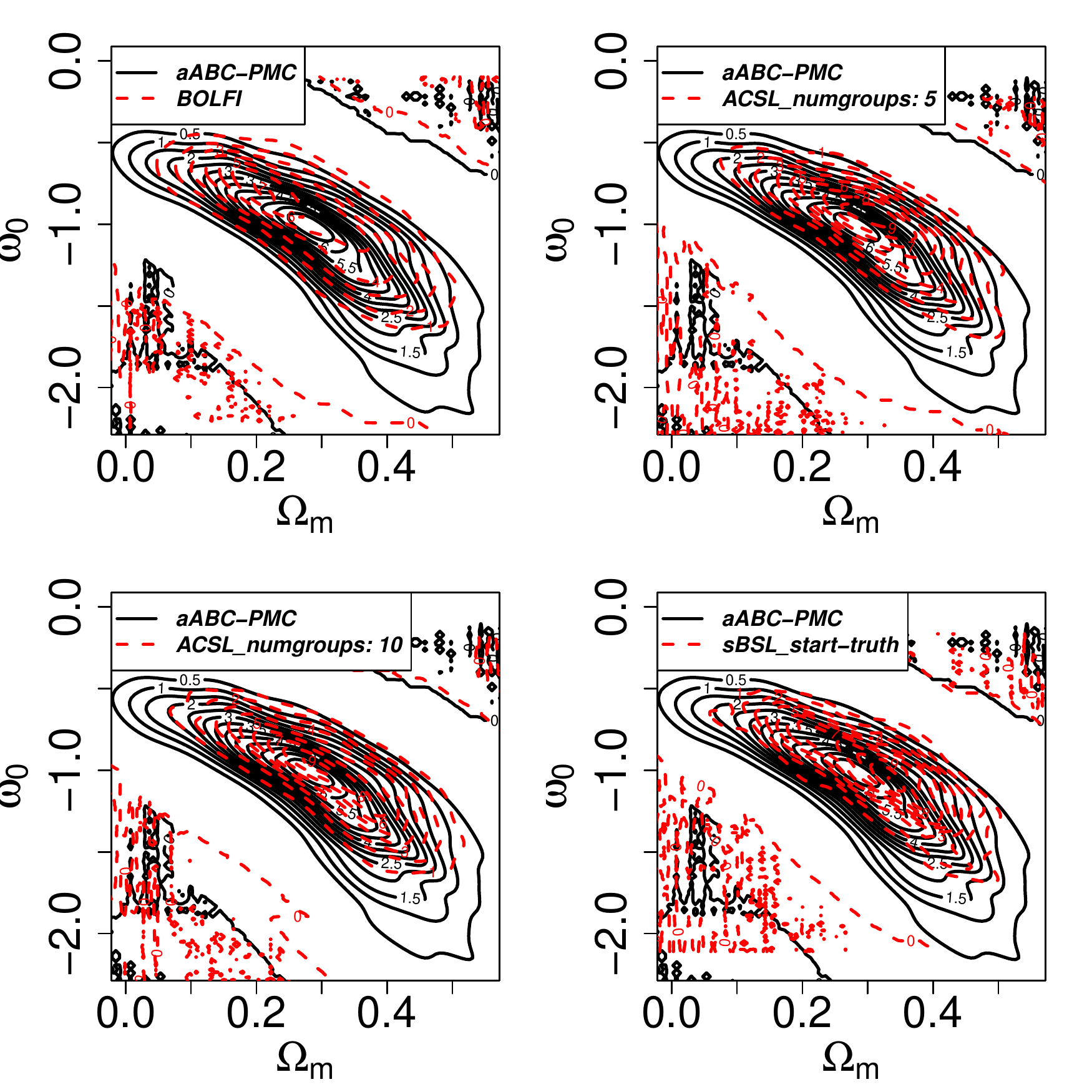}
   \caption{Supernova model. Contour--plot for aABC--PMC method (solid black line), compared with the contour--plots for the remaining methods (dashed red line). In red dashed lines, BOLFI (top left), ACSL with $G=5$ (top right), ACSL with $G=10$ (bottom left) and $\mathrm{sBSL_{truth}}$ (bottom right).
 }
\label{Fig:supernova.sl_contour}
\end{figure}

Traceplots for 11,200 draws from ACSL when $G=5,10$ are in Supplementary Material. Having $G>1$ helps proposals acceptance during the burnin period, so that when ACSL starts it is provided with useful information from the burnin. 
The output of aABC--PMC is produced by 1,000 particles (the final tolerance that is automatically selected by the algorithm after $T=9$ iterations is $\epsilon_9 = 30.5$). For comparison with aABC--PMC inference, where the latter is produced by a ``cloud'' of particles, we thin the output of the single chains of BSL and ACSL: we take the last 10,000 draws from ACSL and sBSL and retain every 10th draw, thus obtaining 1,000 draws that are used to report inference in Table \ref{table.results.mh}. We remind the reader that sBSL fails when initialised at the same starting parameters used for ACSL: therefore to enable some comparison we start sBSL at the ground-truth parameters (this case is denoted $\mathrm{sBSL}_\mathrm{truth}$ in the table). 
Regarding BOLFI, posterior samples were produced by first obtaining $2,000$ ``acquisition points'' in \texttt{ELFI} (over which a GP model is fitted), then 10,000 draws are produced via MCMC, and finally chains were thinned to obtain 1,000 draws used for statistical inference. 
Comparisons between all methods are in Table \ref{table.results.mh} and Figure \ref{Fig:supernova.sl_contour}. Inference results for $\mathrm{sBSL_{truth}}$, ACSL (both attempts) and BOLFI are similar, however the ESS for BOLFI is the highest, while the ESS for $\mathrm{sBSL_{truth}}$ is much lower than for ACSL, as reported in Table \ref{table.results.mh}. While for this case study the exact posterior is unknown, we can speculate that discrepancies in terms of  posterior variability between the results obtained with aABC-PMC and those obtained with the other methods  can likely be explained by the use of the $20$--dimensional summary statistics. For a study on the impact of the summaries dimension in an ABC analysis we refer the reader to \cite{blum2013comparative}.
 As a further remark, it is important to remember that, unlike standard BSL, ACSL was able to get initialised relatively far from ground-truth parameters and still able to return reasonable inference.

\begin{table}[ht]
\begin{center}
\resizebox{\columnwidth}{!}{%
\begin{tabular}{cccccccc}
\hline 
 &  truth & aABC--PMC & $\mathrm{sBSL_{truth}}$ & sBSL  & ACSL, $G=5$ & ACSL, $G=10$ & BOLFI\\
\hline 
$\Omega_m$ & 0.3 &  0.31 (0.071,  0.54)  & 0.313 (0.136, 0.474) & NA &  0.316 (0.133, 0.490) & 0.317 (0.129, 0.488) & 0.289 (0.0765, 0.467)\\ 
$w_0$ & -1 & -1.05 (-1.95, -0.52) &  -1.014 (-1.517, -0.580) & NA & -1.028 (-1.574, -0.607) & -1.047 (-1.502, -0.563) & -0.99 (-1.540, -0.545)\\
minESS & & -- & 301 & NA & 781 & 681 & 831\\
\hline
\end{tabular}
}
\caption{Supernova model: posterior means (95\% HPD interval) resulting from 1,000 thinned posterior draws from several methods. All chains are initialised at $(\Omega_m=0.90, w_0= -0.5)$, except for $\mathrm{sBSL_{truth}}$ which is sBSL initialised at ground-truth parameters. The ``NA'' for sBSL means that the MCMC was unable to move away from the starting location.}
\label{table.results.mh}
\end{center}
\end{table}

\subsection{Simple recruitment, boom and bust with highly skewed summaries}\label{sec:boom-bust}

Here we consider an example that  is discussed in \cite{fasiolo2018extended} and \cite{an2018robust} as it proved challenging due to the  highly non-Gaussian summary statistics. The recruitment boom and bust model is a discrete stochastic temporal model that can be used to represent the fluctuation of the population size of a certain group over time. Given the population size $N_t$ and parameter $\theta=(r,\kappa,\alpha,\beta)$, the next value $N_{t+1}$ has the following distribution
	
	\begin{align*}
		N_{t+1} \sim 
		\begin{cases}
			\mathrm{Poisson}(N_t(1+r)) + \epsilon_t, & \text{ if }\quad N_t \leq \kappa \\
			\mathrm{Binom}(N_t,\alpha) + \epsilon_t, & \text{ if}\quad N_t > \kappa
		\end{cases},
	\end{align*}
	where $\epsilon_t \sim \mathrm{Pois}(\beta)$. The population oscillates between high and low level population sizes for several cycles. Same as in \cite{an2018robust}, true parameters are $r=0.4$, $\kappa=50$, $\alpha=0.09$ and $\beta=0.05$ and we assume $N_1=10$ a fixed and known constant. This value of $\beta$ is considered as it gives rise to highly non-Gaussian summaries, and hence it is of interest to test our methodology in such scenario. In fact, the smaller the value of $\beta$, the more problematic it is to use synthetic likelihoods.
	An illustration of the summaries distribution at the true parameters values is in  Supplementary Material, together with the prior specifications, the summary statistics employed and other model specifications.

\begin{figure}
    \centering
    \begin{subfigure}[b]{0.47\textwidth}
    \includegraphics[width=\textwidth]{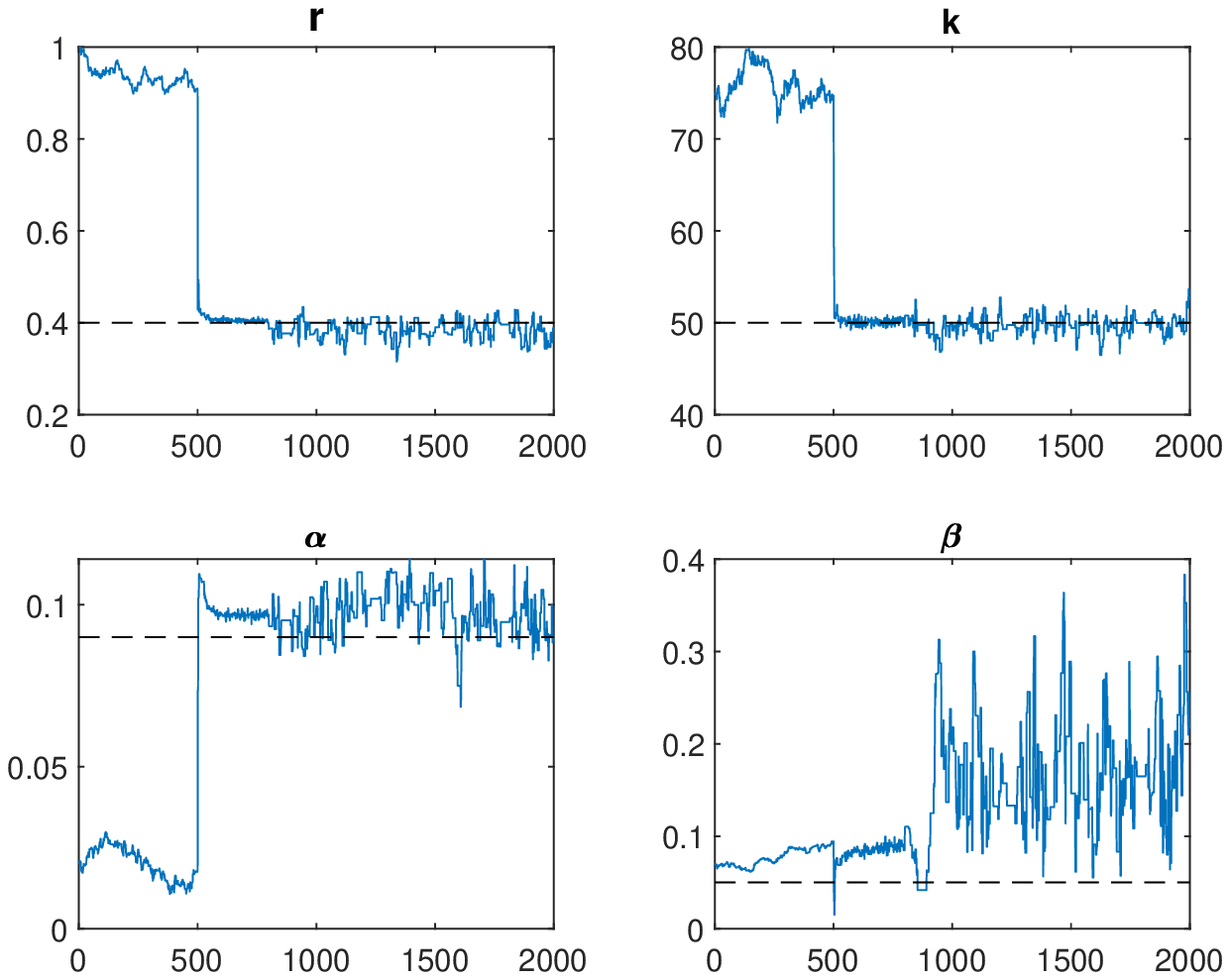}
    \caption{}
    \label{fig:recruitment-asl-set2}
    \end{subfigure}
\begin{subfigure}[b]{0.47\textwidth}
    \includegraphics[width=\textwidth]{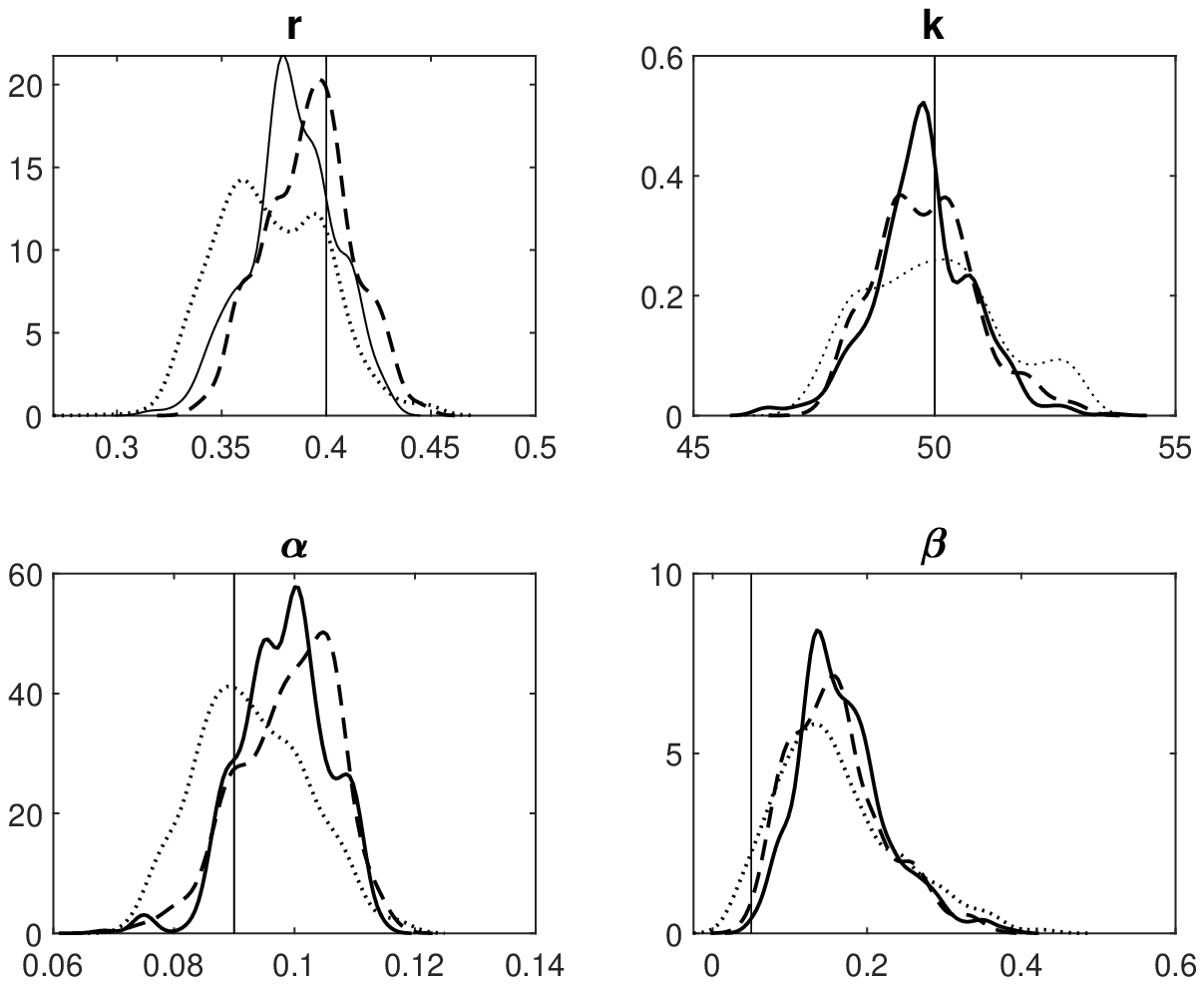}
    \caption{}
    \label{fig:recruitment-asl-bsl-semibsl}
    \end{subfigure}
    \caption{Boom-and-bust: (a) traces for ASL initialised at set 2. Dashed lines are true parameter values. (b) marginal posteriors from 1,000 draws produced with initialisation via ASL (solid) with starting parameters in set 2; with BSL (dashed) and semiBSL (dotted) both initialised at ground truth parameters (vertical lines).}
\end{figure}

We experiment with two sets of values for the starting parameters: set 1 has  $r=0.8$, $\kappa=65$, $\alpha=0.05$,   $\beta=0.07$; set 2 has a more extreme set of values, given by $r=1$, $\kappa=75$, $\alpha=0.02$,   $\beta=0.07$. We always use $M=200$ (also considered in \citealp{an2018robust}). In this case-study we could not experiment with the correlated synthetic likelihoods approach, since the state-of-art generation of Poisson draws requires executing a \texttt{while-loop}, where uniform draws are simulated at each iteration. Therefore it is not known in advance how many uniform draws it is necessary to store, and the implementation of correlated SL results inconvenient. When parameters are initialised in set 1, a burnin of 200 iterations aided by MCWM is considered (MCWM is not used after burnin). When initialising from set 2, we use a longer burnin of 500 iterations. During the burnin, as usual we propose parameters using a Gaussian random walk proposal with constant diagonal covariance matrix with diagonal elements $[0.005^2, 0.5^2, 0.001^2, 0.001^2]$. For ASL the burnin was followed by 300 iterations (again this can be set much smaller) using the guided proposals approach, and then further 1,200 iterations using ``Haario''. BSL was found to diverge to wrong regions of the posterior surface with chains stuck for long periods, for both attempted starting parameters. We therefore implemented the semi-parametric BSL approach from \cite{an2018robust}, thereafter ``semiBSL'': semiBSL is a robustified version of BSL to address the case of non-Gaussian-distributed summary statistics.
However, also semiBSL failed when parameters were initialized in the tails of the posterior (i.e. when using the same starting parameters considered above for ASL), meaning that chains were unable to mix, and were stuck in wrong regions, see the Supplementary Material for details. This shows that even a ``robustified'' version of synthetic likelihoods can be fragile to bad initializations. Therefore, results we report in Figure \ref{fig:recruitment-asl-bsl-semibsl} for both standard BSL and semiBSL are based on chains initialized at the ground-truth parameter values.
With ASL, at the end of the initial 500 burnin iterations we notice the characteristic ``jump'' towards the true parameter values, see Figure \ref{fig:recruitment-asl-set2}.
Therefore, ASL is able to produce inference also when initialised at parameters in the tails of the posterior surface, while BSL and semiBSL cannot, at least for this example. Traces for the failing semiBSL initialised at set 2 are in the Supplementary material. Similarly to the supernova model, we now compute minESS values. Using 1,000 posterior draws from the run initialised at set 2, we have that with ASL minESS is 49. For BSL, when starting at ground-truth, we have minESS=35 and for semiBSL minESS=41. Therefore, the values for semiBSL and ASL are quite similar, despite the fact that ASL is initialised at a less favorable location.

\subsection{Exploration of a multimodal surface}\label{sec:mixture}

\begin{figure}
    \centering
    \begin{subfigure}[b]{0.4\textwidth}
    \includegraphics[width=\textwidth]{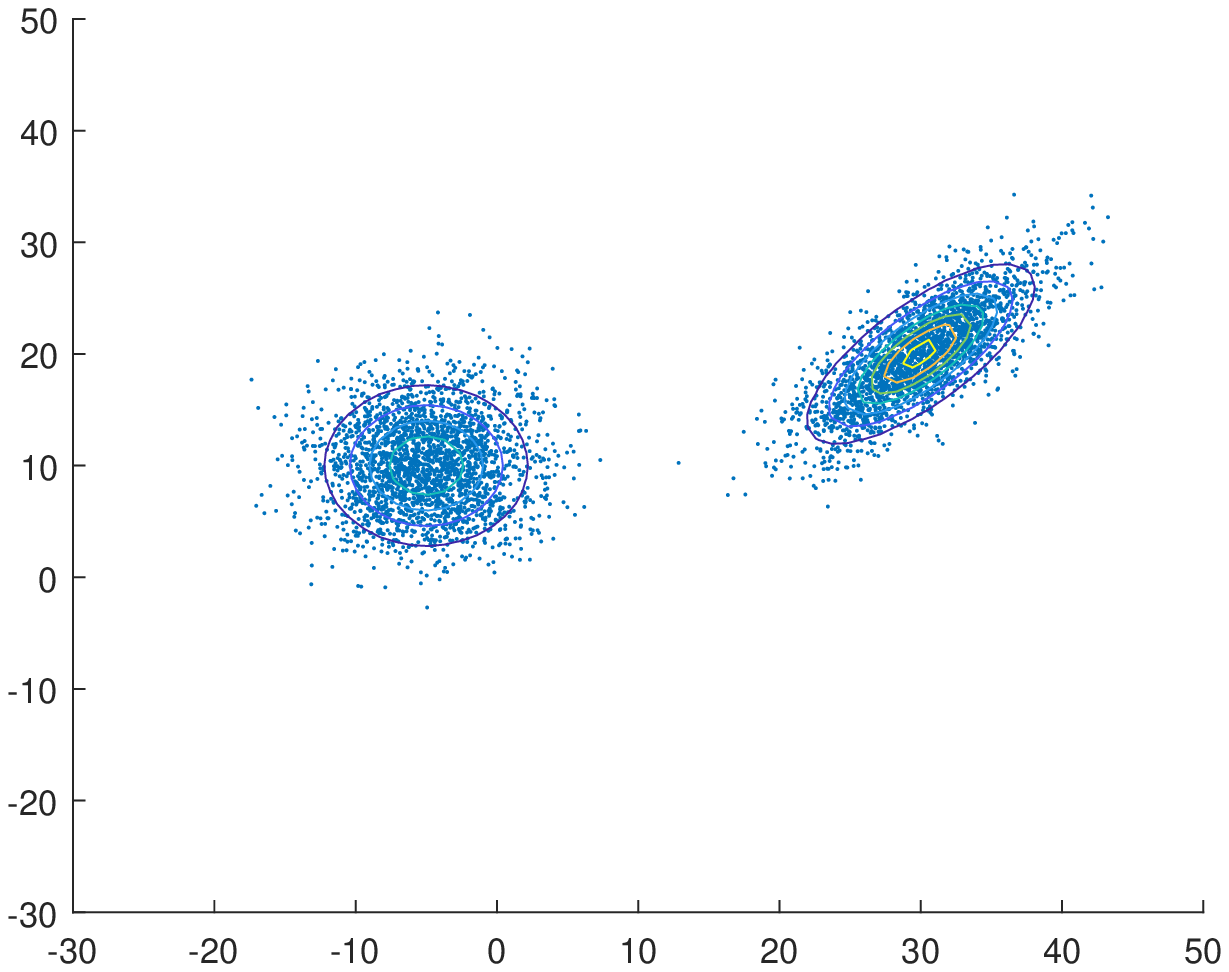}
    \caption{}
    \end{subfigure}
    \begin{subfigure}[b]{0.4\textwidth}
    \includegraphics[width=\textwidth]{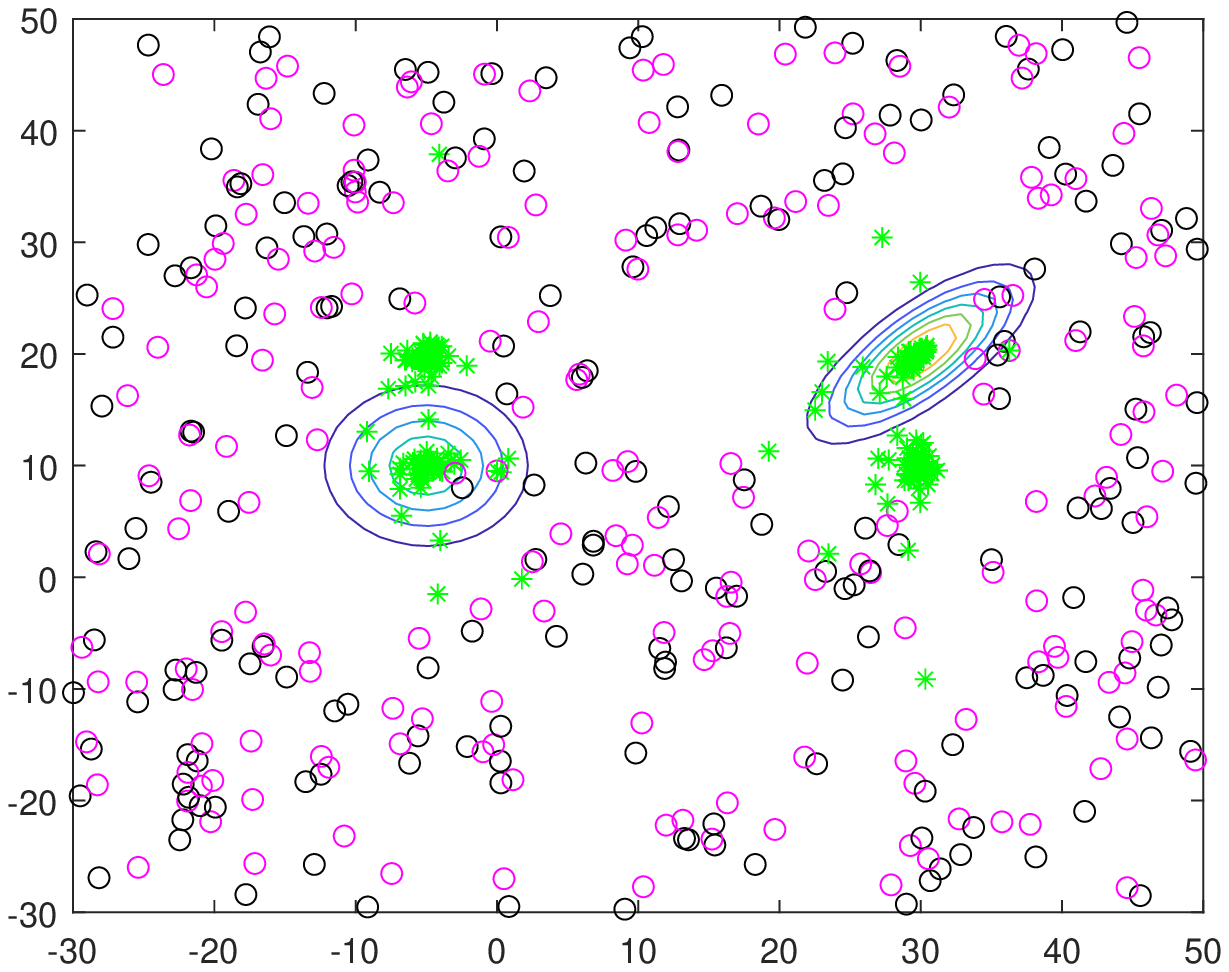}
    \caption{}
    \end{subfigure}
    \caption{Gaussian-mixture: (a)  5,000 data-points from the likelihood model and contour lines for the latter; (b) contour lines for the likelihood model; black circles are the 100 starting values for the corresponding 100 chains; magenta circles correspond to iteration 49 (last burnin iteration) for each chain; the green asterisks correspond to iteration 50 for each chain, that is the first ASL iteration.}
    \label{fig:gm}
\end{figure}

We show how to run multiple chains employing ASL to rapidly alert the researcher of the existence of multiple modes, at a small computational cost. The toy model is admittedly very simple, but the experiment is expressive enough for our take-home message. We consider a likelihood consisting of a two-components Gaussian mixture $x\sim 0.5\mathcal{N}(\mu_1,\Sigma_1)+0.5\mathcal{N}(\mu_2,\Sigma_2)$, where each component is two-dimensional. We consider 5,000 observations generated by such mixture with ground truth $\mu_1 = (\mu_1^{(1)},\mu_1^{(2)}) = (-5,10)$, $\mu_2 = (\mu_2^{(1)},\mu_2^{(2)})= (30,20)$ and covariance matrices $\Sigma_1$ and $\Sigma_2$ both having diagonal entries $(4^2,4^2)$, however $\Sigma_1$ is diagonal while $\Sigma_2$ has off-diagonal entries both equal to 12. Data are exemplified in Figure \ref{fig:gm}(a). We assume $\mu_1$ and $\mu_2$ as the only unknowns, and everything else is fixed to ground-truth values. We set independent priors  $\mu_1^{(1)}\sim \mathcal{N}(-5,2^2)$, $\mu_1^{(2)}\sim \mathcal{N}(10,2^2)$, $\mu_2^{(1)}\sim \mathcal{N}(30,2^2)$, $\mu_2^{(2)}\sim \mathcal{N}(20,2^2)$. In our experiments, summary statistics of simulated data are the estimated means of the two mixture components, as obtained by fitting a two-components Gaussian mixture (with known covariances set to ground-truth). Observed summaries are always $s=(-5,10,30,20)$, that is the ground truth means. We used common strategies to get around the well-known ``label-switching'' issue affecting mixture models: that is  whenever during MCMC a vector $(\mu_1^{(1)},\mu_1^{(2)},\mu_2^{(1)},\mu_2^{(2)})$ is proposed, we sort its entries across the mixture components so that the proposed vector has components rearranged to have $\mu_1^{(1)}<\mu_2^{(1)}$ and $\mu_1^{(2)}<\mu_2^{(2)}$. Since we work in the context of synthetic likelihoods, once a simulated dataset is produced at the proposed parameters, we fit a two-components Gaussian mixture to the data as previously mentioned, and the four corresponding estimated means (which are used as summary statistics) are sorted in the same way as the proposed parameters. 

We use $M=10$ to approximate the synthetic likelihood and design the following experiment. For a fixed dataset with observed summaries $s=(-5,10,30,20)$ we run 100 independent chains initialised at random locations. We set up a very short burnin consisting of 49 iterations where as usual MCWM is used and a Gaussian random walk sampler is employed, where the noise in the random walk has standard deviation set to 0.2 for each proposed entry in $\mu_1$ and $\mu_2$. This means that during burnin we intentionally induce slow exploration of the posterior surface. We show that, as soon as ASL starts, most chains quickly reach the high-density region of the posterior. Figure \ref{fig:gm}(b)  shows 100 starting values that were randomly sampled uniformly in the 4-dimensional hypercube $[-30,50]^4$. We notice that after 49 iterations using random walk proposals the draws are still fairly close to the starting value, however one further iteration afterwards, when ASL is initialised, a rapid jump is performed towards the high density region. The clustering of the ASL draws should signal the researcher the existence of more than one mode, and hence inform her of the opportunity to initialise more than one chain for a full-fledged Bayesian inference, by picking the starting values in the clusters determined by ASL. 

\section{Discussion}

We have introduced several ways to improve the performance of the computing-intensive synthetic likelihood framework. Firstly, we have developed a sequential strategy to learn a ``guided by data'' proposal distribution for SL. The resulting sequentially adaptive and guided SL sampler (ASL) helped the chain to rapidly approach the ground truth parameter values. Importantly, for two of the considered case studies (supernova cosmological parameters and recruitment boom-and-bust model), standard SL methods failed when initialized at remote parameter values and when the standard adaptive MCMC strategy by \cite{haario2001adaptive} was employed, whereas ASL helped the chains to rapidly converge to high-posterior regions (remarkably, this happened even for the markedly non-Gaussian summary statistics considered in section \ref{sec:boom-bust}).
In addition, we have shown how to introduce correlation between successive estimates of the synthetic likelihood, calling this approach ``correlated synthetic likelihoods''. This should help reducing the variance in the acceptance ratio of Metropolis-Hastings, and indeed we have noticed an increase in the mixing of the chains. We have shown how this correlated SL approach (CSL) can be of help when SL is initialized in the tails of the posterior and how beneficial CSL is in terms of chains mixing. However, CSL is not a silver bullet, and it does not necessarily have to succeed at completely eliminating the possibility for SL getting stuck when badly initialized. However, when it can be implemented, there is no obvious reason to prefer standard SL to CSL. At worst, we conjecture that for very nonlinear transformations of the data following the construction of possibly complex summary statistics (and hence complex transformations of the pseudo-random variates), it may happen that the correlation between successive likelihoods gets destroyed, thus transforming CSL into standard SL. We have challenged CSL with a ``perturbed $\alpha$-stable model'' (in Supplementary Material) and even in this case CSL has shown beneficial.
Finally, for the g-and-k and supernova examples, we have illustrated how the problem of a difficult initialization for SL can be tackled by using a Bayesian optimization-based approach to likelihood-free inference \citep{gutmann2016bayesian}, available in the \texttt{ELFI} software \citep{lintusaari2018elfi}. However, we note further that the BOLFI implementation uses the LCB (lower confidence bound) acquisition function which can be prone to over-explore boundaries of parameter spaces and may in some cases result in a poorly resolved surrogate model. An improved acquisition function based on expected integrated variance introduced by \cite{eivba} has been shown to lead to more accurate posterior approximation and it is also available in \texttt{ELFI}, although it is typically rather expensive computationally. As a summary, we believe that when a reasonable starting region where to set an initial $\theta$ is unknown, BOLFI can likely much more rapidly screen the posterior surface in the search for a promising starting region than a random walk proposal. On the other hand, when the dimension of $\theta$ is small as it is often the  case in BSL applications (say $\leq 7$ parameters), then our approach of producing a small number of random walk proposals followed by a short run of ASL can also be more computationally convenient and generally easy to implement.

The steps taken in this work thus broaden the scope of usage of synthetic likelihood methods and open up new venues for further research on improving applicability of intractable inference.

\section{Acknowledgments}

We would like to thank Christopher Drovandi for  useful  feedback  on  an  earlier  draft  of  this  paper. We also thank the anonymous reviewers for helping improve the paper, and especially the reviewer who suggested storing the sample mean of simulated summaries $\bar{s}^{*}$ into $\mathcal{D}$.  UP is supported by the Swedish Research Council (Vetenskapsrådet 2019-03924) and the Chalmers AI Research Centre (CHAIR). JC was funded by the ERC grant no. 742158. US was funded by Academy of Finland grant no. 320182.

\bibliographystyle{abbrvnat}
\bibliography{biblio}

\begin{thebibliography}{73}
\providecommand{\natexlab}[1]{#1}
\providecommand{\url}[1]{\texttt{#1}}
\expandafter\ifx\csname urlstyle\endcsname\relax
  \providecommand{\doi}[1]{doi: #1}\else
  \providecommand{\doi}{doi: \begingroup \urlstyle{rm}\Url}\fi

\bibitem[Allingham et~al.(2009)Allingham, King, and
  Mengersen]{allingham2009bayesian}
D.~Allingham, R.~King, and K.~Mengersen.
\newblock Bayesian estimation of quantile distributions.
\newblock \emph{Statistics and Computing}, 19\penalty0 (2):\penalty0 189--201,
  2009.

\bibitem[An et~al.(2020)An, Nott, and Drovandi]{an2018robust}
Z.~An, D.~J. Nott, and C.~Drovandi.
\newblock Robust bayesian synthetic likelihood via a semi-parametric approach.
\newblock \emph{Statistics and Computing}, 30:\penalty0 543–557, 2020.

\bibitem[Andrieu and Thoms(2008)]{andrieu2008tutorial}
C.~Andrieu and J.~Thoms.
\newblock A tutorial on adaptive {MCMC}.
\newblock \emph{Statistics and computing}, 18\penalty0 (4):\penalty0 343--373,
  2008.

\bibitem[Andrieu et~al.(2009)Andrieu, Roberts, et~al.]{andrieu2009pseudo}
C.~Andrieu, G.~O. Roberts, et~al.
\newblock The pseudo-marginal approach for efficient {M}onte {C}arlo
  computations.
\newblock \emph{The Annals of Statistics}, 37\penalty0 (2):\penalty0 697--725,
  2009.

\bibitem[Andrieu et~al.(2010)Andrieu, Doucet, and
  Holenstein]{andrieu2010particle}
C.~Andrieu, A.~Doucet, and R.~Holenstein.
\newblock Particle {M}arkov chain {M}onte {C}arlo methods.
\newblock \emph{Journal of the Royal Statistical Society: Series B},
  72\penalty0 (3):\penalty0 269--342, 2010.

\bibitem[Beaumont(2003)]{beaumont2003estimation}
M.~A. Beaumont.
\newblock Estimation of population growth or decline in genetically monitored
  populations.
\newblock \emph{Genetics}, 164\penalty0 (3):\penalty0 1139--1160, 2003.

\bibitem[Beaumont et~al.(2009)Beaumont, Cornuet, Marin, and
  Robert]{BeaumontEtAl2009}
M.~A. Beaumont, J.-M. Cornuet, J.-M. Marin, and C.~P. Robert.
\newblock Adaptive approximate bayesian computation.
\newblock \emph{Biometrika}, 96\penalty0 (4):\penalty0 983--990, 2009.

\bibitem[Blum et~al.(2013)Blum, Nunes, Prangle, Sisson,
  et~al.]{blum2013comparative}
M.~G. Blum, M.~A. Nunes, D.~Prangle, S.~A. Sisson, et~al.
\newblock A comparative review of dimension reduction methods in approximate
  bayesian computation.
\newblock \emph{Statistical Science}, 28\penalty0 (2):\penalty0 189--208, 2013.

\bibitem[Botha et~al.(2021)Botha, Kohn, and Drovandi]{botha2021particle}
I.~Botha, R.~Kohn, and C.~Drovandi.
\newblock Particle methods for stochastic differential equation mixed effects
  models.
\newblock \emph{Bayesian Analysis}, 16\penalty0 (2):\penalty0 575--609, 2021.

\bibitem[Chambers et~al.(1976)Chambers, Mallows, and Stuck]{chambers1976method}
J.~M. Chambers, C.~L. Mallows, and B.~Stuck.
\newblock A method for simulating stable random variables.
\newblock \emph{Journal of the American Statistical Association}, 71\penalty0
  (354):\penalty0 340--344, 1976.

\bibitem[Cheng and Higham(1998)]{cheng1998modified}
S.~H. Cheng and N.~J. Higham.
\newblock A modified {C}holesky algorithm based on a symmetric indefinite
  factorization.
\newblock \emph{SIAM Journal on Matrix Analysis and Applications}, 19\penalty0
  (4):\penalty0 1097--1110, 1998.

\bibitem[Chevallier and Polarski(2001)]{chevallier2001accelerating}
M.~Chevallier and D.~Polarski.
\newblock Accelerating universes with scaling dark matter.
\newblock \emph{International Journal of Modern Physics D}, 10\penalty0
  (02):\penalty0 213--223, 2001.

\bibitem[Choppala et~al.(2016)Choppala, Gunawan, Chen, Tran, and
  Kohn]{choppala2016bayesian}
P.~Choppala, D.~Gunawan, J.~Chen, M.-N. Tran, and R.~Kohn.
\newblock Bayesian inference for state space models using block and correlated
  pseudo marginal methods.
\newblock \emph{arXiv preprint arXiv:1612.07072}, 2016.

\bibitem[Condon and Matthews(2018)]{condon2018lambdacdm}
J.~Condon and A.~Matthews.
\newblock $\lambda$cdm cosmology for astronomers.
\newblock \emph{Publications of the Astronomical Society of the Pacific},
  130\penalty0 (989):\penalty0 073001, 2018.

\bibitem[Dahlin et~al.(2015)Dahlin, Lindsten, Kronander, and
  Sch{\"o}n]{dahlin2015accelerating}
J.~Dahlin, F.~Lindsten, J.~Kronander, and T.~B. Sch{\"o}n.
\newblock Accelerating pseudo-marginal {M}etropolis-{H}astings by correlating
  auxiliary variables.
\newblock \emph{arXiv preprint arXiv:1511.05483}, 2015.

\bibitem[Dehideniya et~al.(2019)Dehideniya, Overstall, Drovandi, and
  McGree]{dehideniya2019synthetic}
M.~Dehideniya, A.~M. Overstall, C.~C. Drovandi, and J.~M. McGree.
\newblock A synthetic likelihood-based laplace approximation for efficient
  design of biological processes.
\newblock \emph{arXiv preprint arXiv:1903.04168}, 2019.

\bibitem[Del~Moral et~al.(2012)Del~Moral, Doucet, and Jasra]{del2012adaptive}
P.~Del~Moral, A.~Doucet, and A.~Jasra.
\newblock An adaptive sequential monte carlo method for approximate bayesian
  computation.
\newblock \emph{Statistics and Computing}, 22\penalty0 (5):\penalty0
  1009--1020, 2012.

\bibitem[Deligiannidis et~al.(2018)Deligiannidis, Doucet, and
  Pitt]{deligiannidis2015correlated}
G.~Deligiannidis, A.~Doucet, and M.~K. Pitt.
\newblock The correlated pseudo-marginal method.
\newblock \emph{Journal of the Royal Statistical Society: Series B},
  80\penalty0 (5):\penalty0 839--870, 2018.

\bibitem[D'Errico(2015)]{derrico}
J.~D'Errico.
\newblock nearestspd, 2015.
\newblock
  \url{https://www.mathworks.com/matlabcentral/fileexchange/42885-nearestspd},
  MATLAB Central File Exchange. Retrieved October 8, 2021.

\bibitem[Ding(2016)]{ding2016conditional}
P.~Ding.
\newblock On the conditional distribution of the multivariate t distribution.
\newblock \emph{The American Statistician}, 70\penalty0 (3):\penalty0 293--295,
  2016.

\bibitem[Drovandi and Pettitt(2011)]{drovandi2011likelihood}
C.~Drovandi and A.~Pettitt.
\newblock Likelihood-free {B}ayesian estimation of multivariate quantile
  distributions.
\newblock \emph{Computational Statistics \& Data Analysis}, 55\penalty0
  (9):\penalty0 2541--2556, 2011.

\bibitem[Engblom et~al.(2020)Engblom, Eriksson, and
  Widgren]{engblom2019bayesian}
S.~Engblom, R.~Eriksson, and S.~Widgren.
\newblock Bayesian epidemiological modeling over high-resolution network data.
\newblock \emph{Epidemics}, 32, 2020.

\bibitem[Fasiolo and Wood(2014)]{synlik}
M.~Fasiolo and S.~Wood.
\newblock \emph{An introduction to synlik (2014). R package version 0.1.0.},
  2014.

\bibitem[Fasiolo et~al.(2018)Fasiolo, Wood, Hartig, Bravington,
  et~al.]{fasiolo2018extended}
M.~Fasiolo, S.~N. Wood, F.~Hartig, M.~V. Bravington, et~al.
\newblock An extended empirical saddlepoint approximation for intractable
  likelihoods.
\newblock \emph{Electronic Journal of Statistics}, 12\penalty0 (1):\penalty0
  1544--1578, 2018.

\bibitem[Fearnhead and Prangle(2012)]{fearnhead2012constructing}
P.~Fearnhead and D.~Prangle.
\newblock Constructing summary statistics for approximate {B}ayesian
  computation: semi-automatic approximate bayesian computation.
\newblock \emph{Journal of the Royal Statistical Society: Series B},
  74\penalty0 (3):\penalty0 419--474, 2012.

\bibitem[Genovese et~al.(2009)Genovese, Freeman, Wasserman, Nichol, and
  Miller]{genovese2009inference}
C.~R. Genovese, P.~Freeman, L.~Wasserman, R.~C. Nichol, and C.~Miller.
\newblock Inference for the dark energy equation of state using type ia
  supernova data.
\newblock \emph{The Annals of Applied Statistics}, pages 144--178, 2009.

\bibitem[Ghurye et~al.(1969)Ghurye, Olkin, et~al.]{ghurye1969unbiased}
S.~Ghurye, I.~Olkin, et~al.
\newblock Unbiased estimation of some multivariate probability densities and
  related functions.
\newblock \emph{The Annals of Mathematical Statistics}, 40\penalty0
  (4):\penalty0 1261--1271, 1969.

\bibitem[Golightly et~al.(2019)Golightly, Bradley, Lowe, and
  Gillespie]{golightly2018correlated}
A.~Golightly, E.~Bradley, T.~Lowe, and C.~Gillespie.
\newblock Correlated pseudo-marginal schemes for time-discretised stochastic
  kinetic models.
\newblock \emph{Computational Statistics \& Data Analysis}, 136:\penalty0
  92--107, 2019.

\bibitem[Gutmann and Corander(2016)]{gutmann2016bayesian}
M.~U. Gutmann and J.~Corander.
\newblock Bayesian optimization for likelihood-free inference of
  simulator-based statistical models.
\newblock \emph{The Journal of Machine Learning Research}, 17\penalty0
  (1):\penalty0 4256--4302, 2016.

\bibitem[Guy et~al.(2010)Guy, Sullivan, Conley, Regnault, Astier, Balland,
  Basa, Carlberg, Fouchez, Hardin, et~al.]{guy2010supernova}
J.~Guy, M.~Sullivan, A.~Conley, N.~Regnault, P.~Astier, C.~Balland, S.~Basa,
  R.~Carlberg, D.~Fouchez, D.~Hardin, et~al.
\newblock The supernova legacy survey 3-year sample: Type ia supernovae
  photometric distances and cosmological constraints.
\newblock \emph{Astronomy \& Astrophysics}, 523:\penalty0 A7, 2010.

\bibitem[Haario et~al.(2001)Haario, Saksman, and Tamminen]{haario2001adaptive}
H.~Haario, E.~Saksman, and J.~Tamminen.
\newblock An adaptive {M}etropolis algorithm.
\newblock \emph{Bernoulli}, 7\penalty0 (2):\penalty0 223--242, 2001.

\bibitem[Higham(2015)]{modchol}
N.~Higham.
\newblock Modified cholesky factorization, 2015.
\newblock \url{https://github.com/higham/modified-cholesky}.

\bibitem[Higham(1988)]{higham1988computing}
N.~J. Higham.
\newblock Computing a nearest symmetric positive semidefinite matrix.
\newblock \emph{Linear algebra and its applications}, 103:\penalty0 103--118,
  1988.

\bibitem[Hoffman and Gelman(2014)]{hoffman2014no}
M.~D. Hoffman and A.~Gelman.
\newblock The {No-U-turn} sampler: adaptively setting path lengths in
  {H}amiltonian {M}onte {C}arlo.
\newblock \emph{Journal of Machine Learning Research}, 15\penalty0
  (1):\penalty0 1593--1623, 2014.

\bibitem[Huterer and Turner(2001)]{huterer2001probing}
D.~Huterer and M.~S. Turner.
\newblock Probing dark energy: Methods and strategies.
\newblock \emph{Physical Review D}, 64\penalty0 (12):\penalty0 123527, 2001.

\bibitem[J{\"a}rvenp{\"a}{\"a} et~al.(2019)J{\"a}rvenp{\"a}{\"a}, Gutmann,
  Pleska, Vehtari, and Marttinen]{eivba}
M.~J{\"a}rvenp{\"a}{\"a}, M.~U. Gutmann, A.~Pleska, A.~Vehtari, and
  P.~Marttinen.
\newblock Efficient acquisition rules for model-based approximate bayesian
  computation.
\newblock \emph{Bayesian Analysis}, 14\penalty0 (2):\penalty0 595--622, 2019.

\bibitem[J{\"a}rvenp{\"a}{\"a} et~al.(2020)J{\"a}rvenp{\"a}{\"a}, Gutmann,
  Vehtari, and Marttinen]{jarvenpaa2020parallel}
M.~J{\"a}rvenp{\"a}{\"a}, M.~U. Gutmann, A.~Vehtari, and P.~Marttinen.
\newblock Parallel {G}aussian process surrogate {B}ayesian inference with noisy
  likelihood evaluations.
\newblock \emph{Bayesian Analysis}, 2020.
\newblock \doi{10.1214/20-BA1200}.

\bibitem[Jennings and Madigan(2017)]{jennings2017astroabc}
E.~Jennings and M.~Madigan.
\newblock astroabc: an approximate bayesian computation sequential monte carlo
  sampler for cosmological parameter estimation.
\newblock \emph{Astronomy and computing}, 19:\penalty0 16--22, 2017.

\bibitem[Karabatsos and Leisen(2018)]{karabatsos2018approximate}
G.~Karabatsos and F.~Leisen.
\newblock An approximate likelihood perspective on {ABC} methods.
\newblock \emph{Statistics Surveys}, 12:\penalty0 66--104, 2018.

\bibitem[Kessler et~al.(2009)Kessler, Bernstein, Cinabro, Dilday, Frieman, Jha,
  Kuhlmann, Miknaitis, Sako, Taylor, et~al.]{kessler2009snana}
R.~Kessler, J.~P. Bernstein, D.~Cinabro, B.~Dilday, J.~A. Frieman, S.~Jha,
  S.~Kuhlmann, G.~Miknaitis, M.~Sako, M.~Taylor, et~al.
\newblock Snana: A public software package for supernova analysis.
\newblock \emph{Publications of the Astronomical Society of the Pacific},
  121\penalty0 (883):\penalty0 1028, 2009.

\bibitem[Kokko et~al.(2019)Kokko, Remes, Thomas, Pesonen, and
  Corander]{kokko2019pylfire}
J.~Kokko, U.~Remes, O.~Thomas, H.~Pesonen, and J.~Corander.
\newblock {PYLFIRE}: Python implementation of likelihood-free inference by
  ratio estimation.
\newblock \emph{Wellcome Open Research}, 4\penalty0 (197):\penalty0 197, 2019.

\bibitem[Krzanowski(2000)]{krzanowski1988principles}
W.~Krzanowski.
\newblock \emph{Principles of Multivariate Analysis}.
\newblock OUP Oxford, 2000.

\bibitem[Linder(2003)]{linder2003exploring}
E.~V. Linder.
\newblock Exploring the expansion history of the universe.
\newblock \emph{Physical Review Letters}, 90\penalty0 (9):\penalty0 091301,
  2003.

\bibitem[Lintusaari et~al.(2018)Lintusaari, Vuollekoski,
  Kangasr{\"a}{\"a}si{\"o}, Skyt{\'e}n, J{\"a}rvenp{\"a}{\"a}, Gutmann,
  Vehtari, Corander, and Kaski]{lintusaari2018elfi}
J.~Lintusaari, H.~Vuollekoski, A.~Kangasr{\"a}{\"a}si{\"o}, K.~Skyt{\'e}n,
  M.~J{\"a}rvenp{\"a}{\"a}, M.~Gutmann, A.~Vehtari, J.~Corander, and S.~Kaski.
\newblock Elfi: Engine for likelihood-free inference.
\newblock \emph{Journal of Machine Learning Research}, 19\penalty0 (16), 2018.

\bibitem[Marjoram et~al.(2003)Marjoram, Molitor, Plagnol, and
  Tavar{\'e}]{marjoram2003markov}
P.~Marjoram, J.~Molitor, V.~Plagnol, and S.~Tavar{\'e}.
\newblock Markov chain {M}onte {C}arlo without likelihoods.
\newblock \emph{Proceedings of the National Academy of Sciences}, 100\penalty0
  (26):\penalty0 15324--15328, 2003.

\bibitem[McCulloch(1986)]{mcculloch1986simple}
J.~H. McCulloch.
\newblock Simple consistent estimators of stable distribution parameters.
\newblock \emph{Communications in Statistics-Simulation and Computation},
  15\penalty0 (4):\penalty0 1109--1136, 1986.

\bibitem[Nott et~al.(2019)Nott, Drovandi, and Kohn]{nott2019bayesian}
D.~J. Nott, C.~Drovandi, and R.~Kohn.
\newblock Bayesian inference using synthetic likelihood: asymptotics and
  adjustments.
\newblock \emph{arXiv preprint arXiv:1902.04827}, 2019.

\bibitem[Ong et~al.(2018)Ong, Nott, Tran, Sisson, and
  Drovandi]{ong2018likelihood}
V.~M.-H. Ong, D.~J. Nott, M.-N. Tran, S.~A. Sisson, and C.~C. Drovandi.
\newblock Likelihood-free inference in high dimensions with synthetic
  likelihood.
\newblock \emph{Computational Statistics \& Data Analysis}, 128:\penalty0
  271--291, 2018.

\bibitem[Papamakarios et~al.(2019)Papamakarios, Sterratt, and
  Murray]{papamakarios2018sequential}
G.~Papamakarios, D.~C. Sterratt, and I.~Murray.
\newblock Sequential neural likelihood: Fast likelihood-free inference with
  autoregressive flows.
\newblock In K.~Chaudhuri and M.~Sugiyama, editors, \emph{Proceedings of
  Machine Learning Research}, volume~89 of \emph{Proceedings of Machine
  Learning Research}, pages 837--848, 2019.

\bibitem[Peters et~al.(2012)Peters, Sisson, and Fan]{peters2012likelihood}
G.~W. Peters, S.~A. Sisson, and Y.~Fan.
\newblock Likelihood-free {B}ayesian inference for $\alpha$-stable models.
\newblock \emph{Computational Statistics \& Data Analysis}, 56\penalty0
  (11):\penalty0 3743--3756, 2012.

\bibitem[Picchini and Anderson(2017)]{picchini2017approximate}
U.~Picchini and R.~Anderson.
\newblock Approximate maximum likelihood estimation using data-cloning {ABC}.
\newblock \emph{Computational Statistics \& Data Analysis}, 105:\penalty0
  166--183, 2017.

\bibitem[Picchini and Everitt(2019)]{picchini2019stratified}
U.~Picchini and R.~G. Everitt.
\newblock Stratified sampling and bootstrapping for approximate {B}ayesian
  computation.
\newblock \emph{arXiv preprint arXiv:1905.07976}, 2019.

\bibitem[Picchini and Forman(2019)]{picchini2019bayesian}
U.~Picchini and J.~L. Forman.
\newblock Bayesian inference for stochastic differential equation mixed effects
  models of a tumour xenography study.
\newblock \emph{Journal of the Royal Statistical Society: Series C},
  68\penalty0 (4):\penalty0 887--913, 2019.

\bibitem[Prangle(2017)]{gk}
D.~Prangle.
\newblock gk: An {R} package for the g-and-k and generalised g-and-h
  distributions.
\newblock \emph{arXiv:1706.06889}, 2017.

\bibitem[Prangle et~al.(2017)]{prangle2017adapting}
D.~Prangle et~al.
\newblock Adapting the {ABC} distance function.
\newblock \emph{Bayesian Analysis}, 12\penalty0 (1):\penalty0 289--309, 2017.

\bibitem[Price et~al.(2018)Price, Drovandi, Lee, and Nott]{price2018bayesian}
L.~F. Price, C.~C. Drovandi, A.~Lee, and D.~J. Nott.
\newblock Bayesian synthetic likelihood.
\newblock \emph{Journal of Computational and Graphical Statistics}, 27\penalty0
  (1):\penalty0 1--11, 2018.

\bibitem[Rasmussen and Williams(2006)]{rasmussen2004gaussian}
C.~E. Rasmussen and C.~Williams.
\newblock \emph{Gaussian processes in machine learning}.
\newblock The MIT Press, 2006.

\bibitem[Rayner and MacGillivray(2002)]{rayner2002numerical}
G.~D. Rayner and H.~L. MacGillivray.
\newblock Numerical maximum likelihood estimation for the g-and-k and
  generalized g-and-h distributions.
\newblock \emph{Statistics and Computing}, 12\penalty0 (1):\penalty0 57--75,
  2002.

\bibitem[Robert and Casella(2004)]{robert2004monte}
C.~Robert and G.~Casella.
\newblock \emph{Monte {C}arlo statistical methods}.
\newblock Springer Science \& Business Media, 2004.

\bibitem[Sch{\"o}n et~al.(2018)Sch{\"o}n, Svensson, Murray, and
  Lindsten]{schon2018probabilistic}
T.~B. Sch{\"o}n, A.~Svensson, L.~Murray, and F.~Lindsten.
\newblock Probabilistic learning of nonlinear dynamical systems using
  sequential {M}onte {C}arlo.
\newblock \emph{Mechanical Systems and Signal Processing}, 104:\penalty0
  866--883, 2018.

\bibitem[Simola et~al.(2020)Simola, Cisewski-Kehe, Gutmann, Corander,
  et~al.]{simola2020adaptive}
U.~Simola, J.~Cisewski-Kehe, M.~U. Gutmann, J.~Corander, et~al.
\newblock Adaptive approximate bayesian computation tolerance selection.
\newblock \emph{Bayesian Analysis}, 2020.

\bibitem[Sisson and Fan(2011)]{sisson2011likelihood}
S.~A. Sisson and Y.~Fan.
\newblock \emph{Handbook of {M}arkov chain {M}onte {C}arlo}, chapter
  Likelihood-free {MCMC}.
\newblock Chapman \& Hall/CRC, New York., 2011.

\bibitem[Sisson et~al.(2018)Sisson, Fan, and Beaumont]{sisson2018handbook}
S.~A. Sisson, Y.~Fan, and M.~Beaumont.
\newblock \emph{Handbook of Approximate Bayesian Computation}.
\newblock Chapman and Hall/CRC, 2018.

\bibitem[Thomas et~al.(2021)Thomas, Dutta, Corander, Kaski, and
  Gutmann]{thomas2016likelihood}
O.~Thomas, R.~Dutta, J.~Corander, S.~Kaski, and M.~U. Gutmann.
\newblock Likelihood-free inference by ratio estimation.
\newblock \emph{Bayesian Analysis}, 2021.

\bibitem[Tran et~al.(2016)Tran, Kohn, Quiroz, and Villani]{tran2016block}
M.-N. Tran, R.~Kohn, M.~Quiroz, and M.~Villani.
\newblock The block pseudo-marginal sampler.
\newblock \emph{arXiv preprint arXiv:1603.02485}, 2016.

\bibitem[Tripathi et~al.(2017)Tripathi, Sangwan, and Jassal]{tripathi2017dark}
A.~Tripathi, A.~Sangwan, and H.~Jassal.
\newblock Dark energy equation of state parameter and its evolution at low
  redshift.
\newblock \emph{Journal of Cosmology and Astroparticle Physics}, 2017\penalty0
  (06):\penalty0 012, 2017.

\bibitem[Usmani et~al.(2008)Usmani, Ghosh, Mukhopadhyay, Ray, and
  Ray]{usmani2008dark}
A.~Usmani, P.~Ghosh, U.~Mukhopadhyay, P.~Ray, and S.~Ray.
\newblock The dark energy equation of state.
\newblock \emph{Monthly Notices of the Royal Astronomical Society: Letters},
  386\penalty0 (1):\penalty0 L92--L95, 2008.

\bibitem[Vihola(2012)]{vihola2012robust}
M.~Vihola.
\newblock Robust adaptive metropolis algorithm with coerced acceptance rate.
\newblock \emph{Statistics and Computing}, 22\penalty0 (5):\penalty0 997--1008,
  2012.

\bibitem[Warton(2008)]{warton2008penalized}
D.~I. Warton.
\newblock Penalized normal likelihood and ridge regularization of correlation
  and covariance matrices.
\newblock \emph{Journal of the American Statistical Association}, 103\penalty0
  (481):\penalty0 340--349, 2008.

\bibitem[Weron(1996)]{weron1996chambers}
R.~Weron.
\newblock On the chambers-mallows-stuck method for simulating skewed stable
  random variables.
\newblock \emph{Statistics \& probability letters}, 28\penalty0 (2):\penalty0
  165--171, 1996.

\bibitem[Wetterich(2004)]{wetterich2004phenomenological}
C.~Wetterich.
\newblock Phenomenological parameterization of quintessence.
\newblock \emph{Physics Letters B}, 594\penalty0 (1-2):\penalty0 17--22, 2004.

\bibitem[Wiqvist et~al.(2021)Wiqvist, Golightly, McLean, and
  Picchini]{wiqvist2021efficient}
S.~Wiqvist, A.~Golightly, A.~T. McLean, and U.~Picchini.
\newblock Efficient inference for stochastic differential equation
  mixed-effects models using correlated particle pseudo-marginal algorithms.
\newblock \emph{Computational Statistics \& Data Analysis}, 157:\penalty0
  107151, 2021.

\bibitem[Wood(2010)]{wood2010statistical}
S.~N. Wood.
\newblock Statistical inference for noisy nonlinear ecological dynamic systems.
\newblock \emph{Nature}, 466\penalty0 (7310):\penalty0 1102, 2010.

\end{thebibliography}

\clearpage

\newpage

\begin{center}
  \section*{SUPPLEMENTARY MATERIAL}  
\end{center}

\section*{Adaptive MCMC of Haario et al (2001)}

To make our work self-contained, here we give the covariance update formula for the adaptive Gaussian proposal of \cite{haario2001adaptive}, which we use throughout our paper.
Say that the chain is in position $\theta_{r-1}$ at iteration $r-1$. At iteration $r$ we want to propose $\theta_{r}\sim \mathcal{N}(\theta_{r-1},C_r)$, and the previous MCMC ``history'' is given by $(\theta_0,...,\theta_{r-1})$. Then we can take
\[
 C_r =
\begin{cases}
C_\mathrm{init}, \qquad r\leq K    \\
\frac{2.4^2}{\dim(\theta)}\cdot\mathrm{cov}(\theta_0,...,\theta_{r-1})+\frac{2.4^2}{\dim(\theta)}\cdot\epsilon \cdot \mathrm{I}_{\dim(\theta)}, \qquad r>K.
\end{cases}
\]
Here $K$ is the number of burnin iterations and $C_\mathrm{init}$ is a user-provided covariance matrix of dimensions $\dim(\theta)\times \dim(\theta)$. $\mathrm{I}_{\dim(\theta)}$ is the identity matrix of dimensions $\dim(\theta)\times \dim(\theta)$ and $\epsilon$ is a small positive factor (we set it to $10^{-8}$). As for $\mathrm{cov}(\cdot)$ this denotes the empirical covariance matrix applied to the chain's history. Importantly, it is often detrimental to update $C_r$ at every iteration. We typically update it every 30 or every 50 iterations (again using the entire past history when updating), and this still preserves the validity of the algorithm (see remark 3 in \citealp{haario2001adaptive}).

As for our experiments: before running ASL we run a small number of $K$ burnin iterations by proposing via $\theta_{r}\sim \mathcal{N}(\theta_{r-1},C_\mathrm{init})$, $r\leq K$, where we always assume $C_\mathrm{init}$ to be diagonal (details are given in each case study).
Afterwards we run $T$ ASL iterations. Once these have terminated, we take the (post-burnin) accepted draws from ASL, denoted $(\theta_{1},..,\theta_T)$, and  implement the method by \cite{haario2001adaptive} starting at $\theta_T$ and removing the burnin, that is the very first iteration has covariance matrix computed as
\[
\frac{2.4^2}{\dim(\theta)}\cdot\mathrm{cov}(\theta_1,...,\theta_{T})+\frac{2.4^2}{\dim(\theta)}\cdot\epsilon \cdot \mathrm{I}_{\dim(\theta)},
\]
and afterwards the covariance keeps getting updated by incorporating newly accepted draws into the chain's history (and again, we do not update at each iteration). 

\section*{Bayesian synthetic likelihoods}

Here we provide further details regarding BSL, as found in \cite{price2018bayesian}. A BSL procedure samples from the exact posterior $\pi(\theta|s)$ for any $M$ (note that ``exact'' sampling is ensured only if the distribution of ${s}$ is really Gaussian). The key feature exploits the idea underlying the pseudo-marginal
method of \cite{andrieu2009pseudo}, where an
unbiased estimator is used in place of the unknown likelihood function.
\cite{price2018bayesian} noted that plugging-in the estimates
$\hat{\mu}_{M,\theta} $ and $\hat{\Sigma}_{M,\theta} $ into the Gaussian
likelihood  $p(s|\theta)$ results in a biased estimator $p_M(s|\theta)$ of $p(s|\theta)$. They suggest adopting the unbiased estimator of \cite{ghurye1969unbiased}: 

\begin{align}
\hat{p}({s}|{\theta}) &= (2\pi)^{-d_s/2}\frac{c(d_s,M-2)}{c(d_s,M-1)(1-1/M)^{d_s/2}}|(M-1)\hat{\Sigma}_{M,\theta})|^{-(M-d_s-2)/2}\nonumber\\
& \times \biggl\{\psi\biggl((M-1)\hat{{\Sigma}}_{M,\theta} - \frac{({s}-\hat{{\mu}}_{M,\theta})({s}-\hat{{\mu}}_{M,\theta})'}{(1-1/M)} \biggr)\biggr\}^{(M-d_s-3)/2}.
\label{eq:synlik}
\end{align}
Here $\pi$ denotes the mathematical constant (not the prior), $d_s=\dim(s)$, $M$ is assumed to satisfy $M>d_s+3$, and for a square matrix ${A}$ the function $\psi({A})$ is defined as $\psi({A})=|{A}|$ if ${A}$ is positive definite and $\psi({A})=0$ otherwise, where $|{A}|$ is the determinant of ${A}$. Finally $c(k,v)=2^{-kv/2}\pi^{-k(k-1)/4}/\prod_{i=1}^k\Gamma(\frac{1}{2}(v-i+1))$. We can then plug $\hat{p}({s}|{\theta})$ inside algorithm \ref{alg:synlikMCMC} in place of $p_M({s}|{\theta})$ to obtain a chain targeting $\pi(\theta|s)$, again only if $s$ is Gaussian. This is a powerful result, however in practice the value of $M$ \textit{does affect} the numerical results, as a too low value of $M$ can reduce the mixing of the chain, since the variance of $\hat{p}({s}|{\theta})$ increases for decreasing $M$.

\section*{g-and-k model}

\subsection*{g-and-k: BSL with standard adaptive MCMC and ASL initialization}

\begin{figure}[ht]
    \centering
    \includegraphics[scale=0.7]{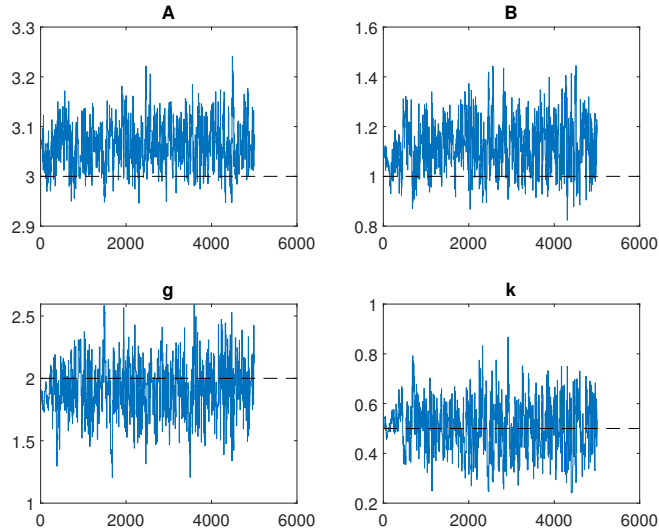}
    \caption{g-and-k: 5,000 iterations of BSL, initialised at the ASL posterior mean, using  $M=50$. The black dashed lines mark ground-truth parameters. }
    \label{fig:g-k_BSL_with_ASL_initialization}
\end{figure}

We consider the case where the adaptive proposal of \cite{haario2001adaptive} is used within BSL and, importantly, the starting parameter value is provided by a chain that used ASL for a few iterations. That is, we consider the draws obtained via ASL from iteration 200 to 500 in Figure \ref{fig:g-and-k_alltraceplots} from the main paper, take their sample mean, and use this to initialize BSL when the ``Haario'' proposal is used. Thanks to the starting value provided by ASL, which is very close to ground truth, we can afford running BSL with only $M=50$. An exemplary run of 5,000 iterations can be obtained in under a minute (with pure Matlab code), with an acceptance rate of around 20\%, see Figure \ref{fig:g-k_BSL_with_ASL_initialization}.

\subsection*{g-and-k: further results using CSL}

\begin{figure}[ht]
    \centering
    \includegraphics[scale=0.7]{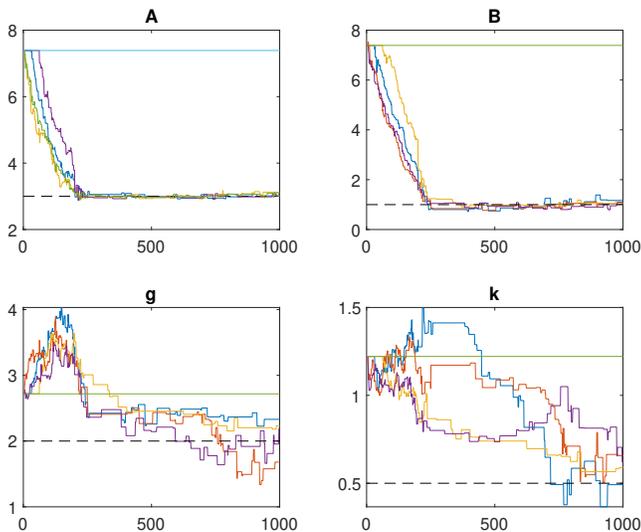}
    \caption{g-and-k: 1,000 iterations of CSL, using  $G=50$ groups. The black dashed lines mark ground-truth parameters. Solid horizontal lines correspond to failed attempts.}
    \label{g-k_CSL_numgroups=50}
\end{figure}

Figure \ref{g-k_CSL_numgroups=50} shows five inference runs for the g-and-k model when CSL is run with $G=50$.

\subsection*{g-and-k: weighting the summaries in BOLFI}

It is possible to assign weights to summary statistics so that the resulting discrepancy is, say, $\Delta=(\sum_{j=1}^{d_s} (s_j^*-s_j)^2/w_j^2)^{1/2}=((s^*-s)'A(s^*-s))^{1/2}$, where $d_s=\dim(s)$. Here the $w_j$ are non-negative weights for each of the components of the summary statistics. Equivalently we may consider the Mahalanobis distance $\Delta=((s^*-s)'A(s^*-s))^{1/2}$, with $A$ interpreted as some scaling matrix (say a covariance matrix). For example we could define $A$ as the diagonal matrix $A=\mathrm{diag}(w_1^{-2},...,w_{d_s}^{-2})$. Summaries are automatically scaled when using the synthetic likelihoods approach (via the $\hat{\Sigma}_M$ matrix), however this is not automatically performed in BOLFI. The reason why it is relevant to give appropriate weights to simulated and observed summaries, is that entries in $s$ and $s^*$ may vary on very different scales, hence $\Delta$ might be dominated by the most variable component of $s$ and $s^*$ (see e.g. \citealp{prangle2017adapting}). Therefore, prior to running BOLFI, we obtain the $w_j$'s in the following way (see also \citealp{picchini2017approximate}). We simulate say $L=1,000$ independent parameter draws from the prior, $\theta^*_l\sim \pi(\theta)$, and simulate corresponding artificial data $y^*_l\sim p(y|\theta^*_l)$, to finally obtain artificial summaries $s^*_l=T(y^*_l)$, $l=1,...,L$. We store all the simulated summaries in a $L\times d_s$ matrix. For each column of this matrix we  compute some robust measure of variability. We consider the median absolute deviation (MAD) as recommended in \cite{prangle2017adapting}, hence obtain $d_s$ MADs, $(\mathrm{MAD}_1,...,\mathrm{MAD}_{d_s})$, and define $w_j:=\mathrm{MAD}_j$, $j=1,...,d_s$. We then construct $A$ as described above, and use BOLFI to optimize $\Delta$. The procedure we have just outlined corresponds to results denoted with \texttt{weighted=yes} in Table \ref{tab:g-and-k_bolfi}. Results using constant $w_j\equiv 1$ are given as \texttt{weighted=no}. The only times we happened to obtain a positive estimate for $k$ was in two instances using weighted summaries. The weighting of summaries statistics is only performed when using BOLFI, not when using the SL approach (in SL, summaries are naturally weighted via the matrix $\hat{\Sigma}$).

\begin{table}[ht]
\small
    \centering
    \begin{tabular}{cllccccc}
    $J_1$ & $J_2$ & weighted & $\min\log\Delta$ &  $\hat{A}$ & $\hat{B}$ & $\hat{g}$ & $\hat{k}$\\
    \hline
    10 & 500 & no   & -0.447 & 3.10 & 1.26 & 1.69 & 0.00\\
    10 & 500 &  yes  & -0.244 & 2.63 & 7-47 & 2.05 & 0.17\\
    20 & 500 & no    &  -0.441 & 3.05 & 1.31 & 1.90 & 0.00\\
    20 & 500 & yes  & -0.194 & 2.90 & 7.83 & 2.2 & 0.00\\
    100 & 500 & no     &  -0.338 &  3.00 & 1.22 & 1.82 & 0.00\\
    100 & 500 & yes & -0.469 & 2.75 & 1.57 & 2.21 & 0.66\\
    200 & 500 & no    &   -0.407 & 3.14 & 1.26 & 1.60  & 0.00\\
    200 & 500 & yes & -0.356 & 3.3 & 0.91 & 2.2 & 0.00\\
    100 & 800 & no & -0.400 & 3.11 & 1.25 & 1.93 & 0.00\\
    100 & 800 & yes & -0.506 & 3.13 & 1.12 & 2.11 & 0.48\\
    \hline
    \end{tabular}
    \caption{g-and-k: values of the optimized log-discrepancies and corresponding parameters for several values of $J_1$ and $J_2$. Ground truth values are $A=3$, $B=1$, $g=2$, $k=0.5$. A ``yes'' in the \texttt{weighted} column implies that discrepancies are computed using weighted summary statistics.}
    \label{tab:g-and-k_bolfi}
\end{table}

\section*{Supernova model}

\subsection*{Simulation from the model}

Here we describe how to simulate a generic dataset. The same procedure is of course used to generate both ``observed data'' and ``simulated data''. We generate $10^4$ variates $u_1,...,u_{10^4}$, independently sampled from a truncated Gaussian $u_j\sim \mathcal{N}_{[0.01,1.2]}(0.5,0.05^2)$ ($j=1,...,10^4$), where $\mathcal{N}_{[a,b]}(m,\sigma^2)$ denotes a Gaussian distribution with mean $m$ and variance $\sigma^2$, truncated to the interval $[a,b]$. The $u_j$ are then binned into 20 intervals of equal width (essentially the bins of an histogram constructed on the $u_j$), then the 20 centres of the bins are obtained and these centres are the ``redshifts'' $z_1,...,z_{20}$. Then for each $z_i$ we compute the distance modulus $\mu_i$ via \eqref{eq:supernovaFM}, using $(\Omega_m,w_0,h_0)=(0.3,-1,0.7)$ ($i=1,...,20$). Therefore, each simulation from the model requires first the generation of the 10,000 truncated Gaussians, then their binning and the calculations of the twenty $\mu_i$. 
Computing the latter is a computational bottleneck, as in order to compute a synthetic likelihood the procedure above has to be performed $M$ times for each new proposed value of $\theta=(\Omega_m,w_0)$.

\subsection*{Supernova model: summaries distribution}

\begin{figure}[htbp]
   \centering
\includegraphics[width=13cm,height=4cm]{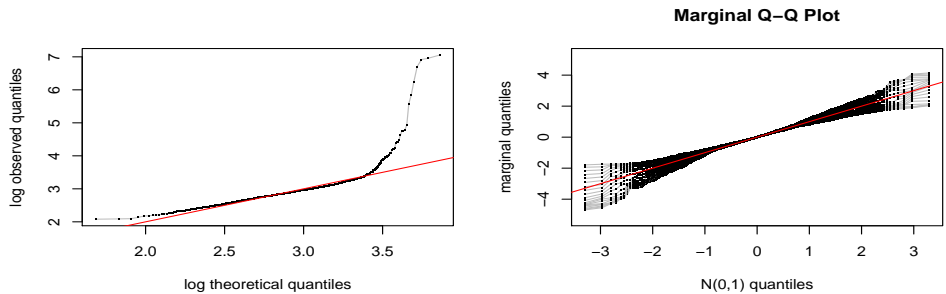} 
   \caption{supernova model: qq-plots for the multivariate summary statistics.}
\label{Fig:supernova.normality.check}
\end{figure}

In order to check if the synthetic likelihood methodology is suitable for conducting the analyses, the multivariate normality assumption of the employed summary statistic must be checked (see \citealp{fasiolo2018extended} and \citealp{an2018robust} for how to relax the assumption). We simulate independently a total of $1,000$ summaries (each having dimension 20), using ground-truth parameters. A test for multivariate normality can be found in \cite{krzanowski1988principles} and is implemented in the \texttt{checkNorm} function from the R package \texttt{synlik} \citep{synlik}, which additionally produces Figure \ref{Fig:supernova.normality.check}. The test does not reject the multivariate normality assumption of the summary statistic at $5\%$ significance level. Furthermore, we note that the right tail behavior in the q-q plot is not unexpected in the synthetic likelihoods context \citep{wood2010statistical}.

\subsection*{Supernova model: general MCMC settings and chains from ACSL}

\begin{figure}[ht]
   \centering
\includegraphics[scale=0.6]{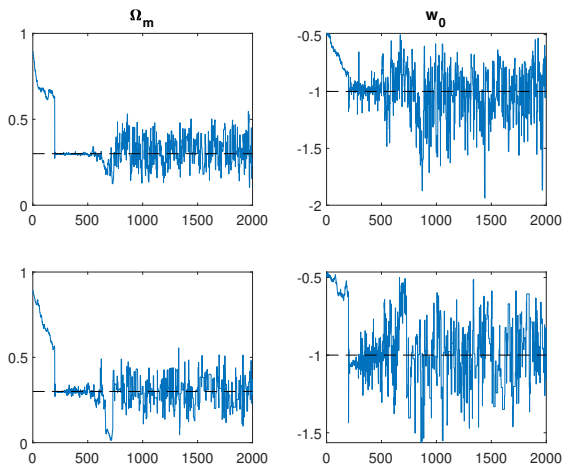} 
   \caption{Supernova model. Trace plots for ACSL corresponding to $G=5$ (top) and $G=10$ (bottom). We show only the first 2,000 non-thinned iterations for ease of display.  Burnin iterations 1--200 use CSL with a Gaussian random walk proposal with constant covariance matrix, while iterations 201--500 use the guided ACSL proposal and the remaining iterations use adaptive MCMC via ``Haario''.}
\label{Fig:supernova.sl_G=5-10}
\end{figure}

For $\mathrm{sBSL_{truth}}$ and ACSL we always have a burnin of $K=200$ iterations, where parameters are proposed using Gaussian random walks, with constant diagonal covariance matrix having standard deviations $[0.01,0.01]$ respectively for $\log\Omega_m$ and $w_0$. For ACSL the burnin is
followed by 300 iterations where parameters are proposed using our guided method, and finally followed by further 10,700 iterations using ``Haario''. The total run therefore comprise $11,200$ iterations. For $\mathrm{sBSL_{truth}}$ the same burnin procedure with constant covariance is used, followed by ``Haario'' for a total of 11,200 iterations. For both $\mathrm{sBSL_{truth}}$ and ACSL the covariance matrix of the summaries used a shrinkage estimator: we considered $\gamma=0.95$ for the shrinkage parameter, which implies a small regularization to  $\hat{\Sigma}_{M,\theta}$, see \cite{nott2019bayesian}.
Figure \ref{Fig:supernova.sl_G=5-10} shows the evolution of the guided and adaptive correlated SL (ACSL) for $G=5,10$: there, only the first 2,000 (out of 11,200) iterations are shown for ease of display. The burnin iterations 1--200 use CSL with a Gaussian random walk proposal with constant covariance matrix, while iterations 201--500 use the guided ACSL proposal and the remaining iterations use adaptive MCMC via \cite{haario2001adaptive}. Having $G>1$ help the chains to move during the burnin period, so that when ACSL starts (iteration 201) it is provided with useful information from the burnin. In fact, it is possible to notice a ``jump'' for $\Omega_m$, which in fact happens at iteration 201, that is when the ACSL kicks in. 

For the interested reader: the overall time required by ELFI/BOLFI was 54 minutes, whereas it took 168 minutes to BSL and 172 minutes to ACSL ($G=10$). Notice this time comparison should be taken with a grain of salt since the latter two methods have an algorithmic structure that is quite different from BOLFI, moreover BOLFI is implemented in Python while ACSL/BSL are implemented in Matlab.

\section*{Recruitment, boom and bust model}

Same as in \cite{fasiolo2018extended} and \cite{an2018robust}, prior distributions are set to $r\sim\mathrm{U}(0,1)$, $\kappa\sim\mathrm{U}(10,80)$, $\alpha\sim\mathrm{U}(0,1)$, $\beta\sim\mathrm{U}(0,1)$. To generate a data set, same as in the cited references we simulate values for the $\{N_t\}$ process for 300 steps, then we discard  the first $50$ values to remove the transient phase of the process. Therefore, data are the remaining $250$ values.  We use essentially the same summary statistics as in \cite{an2018robust}, namely for a dataset $y$, we define differences and ratios as $\mathrm{diff}_{y} = \{y_i - y_{i-1} ; i=2,\ldots,250\}$ and  $\mathrm{ratio}_{y} = \{(y_i+1) / (y_{i-1}+1) ; i=2,\ldots,250\}$, respectively. We use the sample mean, variance, skewness and kurtosis of $y$, $\mathrm{diff}_{y}$ and $\mathrm{ratio}_y$ as our summary statistic, that is a total of twelve summaries. The only difference with the summaries in \cite{an2018robust} is that they take $\mathrm{ratio}_{y} = \{y_i / y_{i-1} ; i=2,\ldots,250\}$, however it is not rare for $\{N_t\}$ to contain zeroes, and  their formulation of $\mathrm{ratio}_{y}$ will cause numerical infelicities.

\begin{figure}
    \centering
    \includegraphics[scale=0.7]{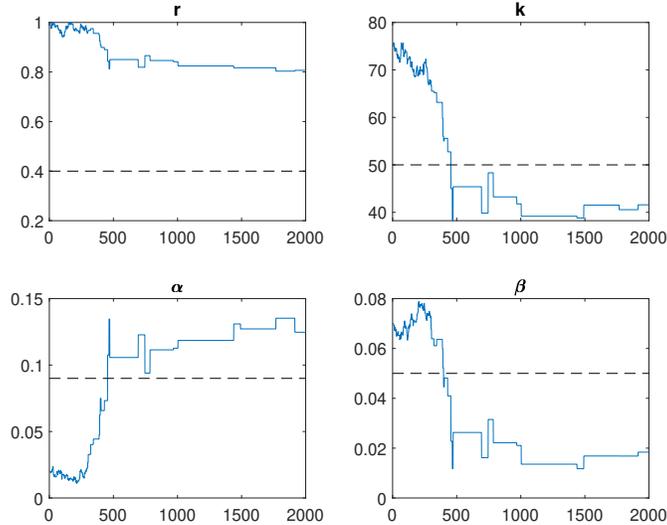}
    \caption{Boom and bust: traces for semiBSL initialised at set 2. Dashed lines are true parameter values.}
    \label{fig:recruitment-semiBSL-set2}
\end{figure}

\begin{figure}
    \centering
    \includegraphics[scale=0.7]{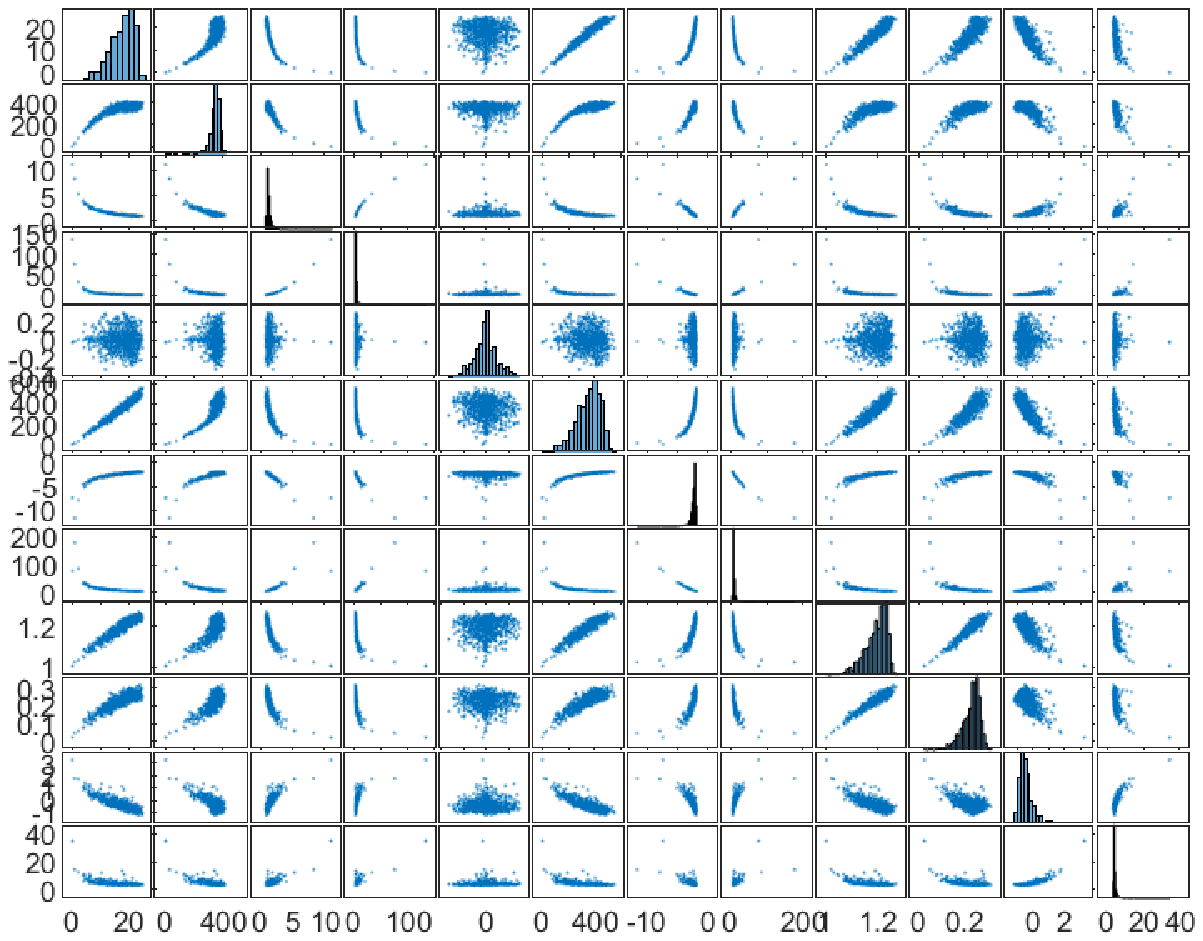}
    \caption{Boom and bust example: scatter plots of 1,000 summaries simulated with $r=0.4$, $\kappa=50$, $\alpha=0.09$ and $\beta=0.05$.}
    \label{fig:recruitment-simsum-scatter}
\end{figure}

As mentioned in the main paper, the boom and bust example is particularly challenging for the BSL approach due to the strong nonlinear dependence structure between the summary statistics. As an illustration,  Figure \ref{fig:recruitment-simsum-scatter} shows the bivariate scatterplots of 1,000 summary statistics simulated with data-generating parameters $r=0.4$, $\kappa=50$, $\alpha=0.09$ and $\beta=0.05$. We initialize parameters at $r=1$, $\kappa=75$, $\alpha=0.02$,   $\beta=0.07$ (this was denoted ``set 2'' in the main paper). We run semiBSL \citep{an2018robust} for 2,000 iterations and use $M=200$, see the main text for full details.  semiBSL  fails to mix properly and ultimately does not converge, see Figure \ref{fig:recruitment-semiBSL-set2}, even though the burnin uses Markov-chain-within-Metropolis (MCWM) to ease mixing. As documented in previous literature, including \cite{an2018robust}, synthetic likelihood approaches can be fragile to bad initializations, but ASL does manage to rapidly converge and mix well, as displayed in the main text.

\section*{$\alpha$-stable and ``perturbed'' $\alpha$-stable distributions}

In this section we wish to challenge CSL by potentially disrupting the positive effect induced by correlating the synthetic loglikelihoods. In order to pursue this goal, we first introduce a sampling algorithm for the $\alpha$-stable distribution, and then perturb this, as explained later.

\subsection*{Sampling from an $\alpha$-stable distribution}

Denote with $\mathrm{Exp}(\lambda)$ the exponential distribution with parameter $\lambda$, then steps 1--4 below illustrate the simulation of a single draw $y$ from an $\alpha$-stable distribution with parameters
$\left(  \alpha,\beta,\gamma,\delta\right)$ \citep{chambers1976method,weron1996chambers}. Notice, while we are not experts of the theoretical derivation of the algorithm below, it is evident that for the case $\alpha=1$ many sources report the implementation of the formula from the highly cited paper \cite{chambers1976method} which, however, was found to contain a typo according to \cite{weron1996chambers} (for $\alpha=1$ \citealp{chambers1976method} has the term $(\pi/2)w\cos(u_2)/(\pi/2+\beta u_2)
$ under the logarithm instead of $w\cos(u_2)/(\pi/2+\beta u_2)
$). We follow \cite{weron1996chambers}.

\begin{enumerate}
\item sample $w \sim \mbox{Exp}(1)$: this can be performed by first simulating $u_1\sim \mbox{Uniform}(0,1)$ and then obtaining $w=-\log(u_1)$;

\item sample $u_2 \sim \mbox{Uniform}(-\pi/2,\pi/2)$;

\item obtain a sample $\tilde{y}$
\[
\tilde{y}=\left\{
\begin{array}
[c]{cll}
&S_{\alpha,\beta}\frac{\sin(\alpha\left(
u_2+B_{\alpha,\beta})\right) }{(\cos u_2)^{1/\alpha}}\left[
\frac{\cos\left(  u_2-\alpha\left( u_2+B_{\alpha,\beta}\right)
\right)  }{w}\right]  ^{\frac{1-\alpha
}{\alpha}}&\text{ \ \ if }\alpha\neq1\\
&\frac{2}{\pi}\left[  \left(  \frac{\pi}{2}+\beta u_2\right) \tan
u_2-\beta\log\bigl(\frac{w\cos u_2}{\frac{\pi}{2}+\beta u_2
}\bigr)\right] & \text{ \ \ if }\alpha=1
\end{array}
\right.
\]
with $S_{\alpha,\beta}=\left(  1+\beta^{2}\tan^{2}\left(  \frac{\pi\alpha}
{2}\right)  \right)  ^{1/2\alpha}$ and
$B_{\alpha,\beta}=\frac {1}{\alpha}\arctan\left(  \beta\tan\left(
\frac{\pi\alpha}{2}\right) \right) .$ In this case
$\tilde{y}$ will be $\alpha$-stable distributed with parameters
$\left(  \alpha,\beta,1,0\right)  $.

\item Apply 
\[
y=\left\{
\begin{array}
[c]{cll}
&\gamma\tilde{y}+\delta&\text{ \ \ if }\alpha\neq1\\
&\gamma\tilde{y} + \frac{2}{\pi} \beta\gamma \log(\gamma) + \delta& \text{ \ \ if }\alpha=1
\end{array}
\right.
\]
then $y$ will be $\alpha$-stable distributed with parameters
$\left(  \alpha,\beta,\gamma,\delta\right)  $.
\end{enumerate}

\subsection*{A perturbed $\alpha$-stable model}

$\alpha$-stable distributions are utilized as models for heavy-tailed noise in many areas of statistics, finance and signal processing engineering. The univariate $\alpha$-stable distribution is typically specified by four parameters: $\alpha\in (0,2]$ determining the rate of tail decay; $\beta\in[-1,1]$ determining the degree and sign of asymmetry (skewness); $\gamma>0$ the scale (under some parameterizations); and the location $\delta$. In general, $\alpha$-stable models admit no closed form expression for the density, except for the Gaussian ($\alpha=2,\beta=0$) and Cauchy ($\alpha=1,\beta=0$)  families, however sampling is straightforward.  Properties of the $\alpha$-stable distributions are widely available, and the interested reader can refer for example to \cite{peters2012likelihood} (also discussing the multivariate case). We only study the benefits induced by using correlated synthetic likelihoods (CSL) for a ``perturbed'' version of the univariate $\alpha$-stable distribution. The perturbed model is defined in a way that, even though in CSL the two synthetic likelihoods corresponding to two consecutive iterations use correlated pseudorandom draws, however the generative models used in the two iterations will be made structurally different. This is supposed to challenge the beneficial effect of using correlated likelihoods, since the correlated draws enter different generative models. As we show, also in this case CSL provides considerably increased mixing.

The procedure for sampling a single draw from a (non-perturbed) $\alpha$-stable distribution was given in the previous section and shows a discontinuity at $\alpha=1$. We use this opportunity to challenge the (potential) benefits brought-in by correlated synthetic likelihoods (CSL): we define a ``perturbed'' version of the (standard) $\alpha$-stable sampler by applying the equations for $\alpha=1$ to the case where $\alpha$ is ``very close to 1'', namely $\alpha \in [0.97,1.03]$, and thereby writing (and everything else staying the same) 
\begin{numcases}
{\tilde{y}=}
S_{\alpha,\beta}\frac{\sin(\alpha\left(
u_2+B_{\alpha,\beta})\right) }{(\cos u_2)^{1/\alpha}}\left[
\frac{\cos\left(  u_2-\alpha\left( u_2+B_{\alpha,\beta}\right)
\right)}{w}\right]^{\frac{1-\alpha}{\alpha}}&\text{if } $\alpha \notin [0.97,1.03]$ \label{eq:perturbed-tilde_1}\\
\frac{2}{\pi}\left[  \left(  \frac{\pi}{2}+\beta u_2\right) \tan
u_2-\beta\ln\frac{\frac{\pi}{2}w\cos u_2}{\frac{\pi}{2}+\beta u_2
}\right] & \text{if } $\alpha \in [0.97,1.03]$\label{eq:perturbed-tilde_2}
\end{numcases}
\begin{numcases}
{{y}=}
\gamma\tilde{y}+\delta&\text{if }$\alpha \notin [0.97,1.03]$\label{eq:perturbed_1}\\
\gamma\tilde{y} + \frac{2}{\pi} \beta\gamma \log(\gamma) + \delta& \text{if }$\alpha \in [0.97,1.03]$\label{eq:perturbed_2},
\end{numcases}
and then we say that the $y$ generated via \eqref{eq:perturbed_1}-\eqref{eq:perturbed_2} is a draw from a perturbed $\alpha$-stable distribution.
This way, depending on the proposed values of $\alpha$ simulated during two consecutive iterations of CSL, the corresponding simulated data may not necessarily use the same equation (e.g. the pair \eqref{eq:perturbed-tilde_1} and \eqref{eq:perturbed_1} as opposed to the pair \eqref{eq:perturbed-tilde_2} and \eqref{eq:perturbed_2}) from the given simulator. As an example, say that at iteration $t$ we accepted an $\alpha$ which had a value smaller than 0.97 or larger than 1.03, and hence accepted a synthetic likelihood that was constructed on simulations from \eqref{eq:perturbed_1}; however at iteration $t+1$ we may propose (and maybe accept) a synthetic likelihood produced from data simulated using some $\alpha\in [0.97,1.03]$ and therefore using \eqref{eq:perturbed_2}. With CSL the two synthetic likelihoods are based on correlated pseudorandom draws, however the generative models used in the two iterations of the example above would be different, thus challenging the beneficial effect of using correlated likelihoods, also because during the inference runs the proposed values of $\alpha$ frequently move inside and outside the region $[0.97,1.03]$. Notice this challenge would not be possible with the non-perturbed $\alpha$-stable model, as in such case, due to floating-point numerical representations, with zero probability a proposed value of $\alpha$ is exactly equal to 1, and hence at each iteration we would always generate from the equation having $\alpha\neq 1$ in the previous section, and this would not challenge CSL.

Next section describes how BSL and CSL are implemented for the perturbed $\alpha$-stable model as well as inference results.

\subsection*{Inference via BSL and CSL for perturbed $\alpha$-stable distributions}

Notice inference for $\alpha$-stable distributions using synthetic likelihoods has been considered in \cite{ong2018likelihood}, and inference is conducted so that we first initialise transformed parameters $(\tilde{\alpha},\tilde{\beta},\tilde{\gamma},\tilde{\delta})$ defined as 
\begin{align}
\tilde{\alpha} = \frac{\log(\alpha-0.5)}{2-\alpha},\quad \tilde{\beta}=\frac{\log(\beta+1)}{1-\beta},\quad \tilde{\gamma}=\log(\gamma),\quad \tilde{\delta}=\delta
\end{align}
so that, when transformed back on the original scale (inside the model simulator), we are ensured that $\alpha\in (0.5,2]$, $\beta\in [-1,1]$ and $\gamma>0$.
We set priors on $(\tilde{\alpha},\tilde{\beta},\tilde{\gamma},\tilde{\delta})$ to be independent standard Gaussians for each parameter. The MCMC is therefore advanced on transformed parameters, however our inference results are shown for the original ones. As for the summary statistics, same as in \cite{peters2012likelihood} and \cite{ong2018likelihood}, we consider  a point estimator of $({\alpha},{\beta},{\gamma})$ due to \cite{mcculloch1986simple}, defined as $(S_\alpha,S_\beta,S_\gamma)$ where
\[
S_\alpha = \frac{q_{95}(\cdot) - q_5(\cdot)}{q_{75}(\cdot)-q_{25}(\cdot)},\quad
S_\beta  = \frac{q_{95}(\cdot) + q_5(\cdot) -2q_{50}(\cdot)}{q_{95}(\cdot) - q_5(\cdot)},\quad
S_\gamma = \frac{q_{75}(\cdot)-q_{25}(\cdot)}{\gamma_\mathrm{true}}
\]
with the addition of $S_\delta = \bar{y}$, the sample mean of the data. Here
$q_i(\cdot)$ is the $i$-th empirical percentile of the submitted dataset. Notice $S_\gamma$ is more generally depending on the parameter $\gamma$ which we decided to fix (but only for the purpose of computing the summary statistic $S_\gamma$) to its true value $\gamma_\mathrm{true}=1$ (see below). Notice \textit{we do infer} $\gamma$ in our procedure, except for plugging its true value in $S_\gamma$ for ease of computation (however when dealing with real data, where ground-truth is of course unavailable, \citealp{peters2012likelihood} suggest using the tabled-estimate of $\gamma$ provided in \citealp{mcculloch1986simple} in place of $\gamma_\mathrm{true}$).

When implementing correlated synthetic likelihoods (CSL) with $M$ model simulations, we initialise a matrix of size $n\times 2M$ (where $n$ is the length of the data) where each row contains a pair of pseudo-random uniforms $(u_1,u_2)$. Then at each iteration we induce correlation in the synthetic likelihoods by randomly picking a block of several consecutive columns (how many columns it depends on the value of $G$) in said matrix and sample new pseudo-random uniforms for that block, while keeping the uniform draws in the remaining columns as fixed.

From the perturbed $\alpha$-stable model we simulate an ``observed'' dataset consisting of 500 independent observations using $\alpha=1.01$, $\beta = 0.5$, $\gamma=1$, $\delta=0$. During the inference we exclusively generate from the perturbed model, and the chain for $\alpha$ often transverses the two regimes (i.e. $\alpha \in [0.97,1.03]$ and $\alpha \notin [0.97,1.03]$). We set starting (transformed) parameter values to be $(\tilde{\alpha},\tilde{\beta},\tilde{\gamma},\tilde{\delta})=(0.8,0.95,2,1)$ and for both BSL and CSL we run 5,200 iterations, using the perturbed model, where the first 200 are burnin iterations. When BSL is used, during the burnin MCWM is employed (and never after the burnin) and afterwards we use the ``Haario'' adaptive proposal. With CSL, we never use MCWM and during burnin we use a fixed covariance matrix (and of course we correlate the likelihoods), and use ``Haario'' afterwards. Therefore here we do not use our guided ASL, to focus instead on the benefits of using CSL, which are evident as discussed below.

\begin{figure}
    \centering
    \includegraphics[scale=0.9]{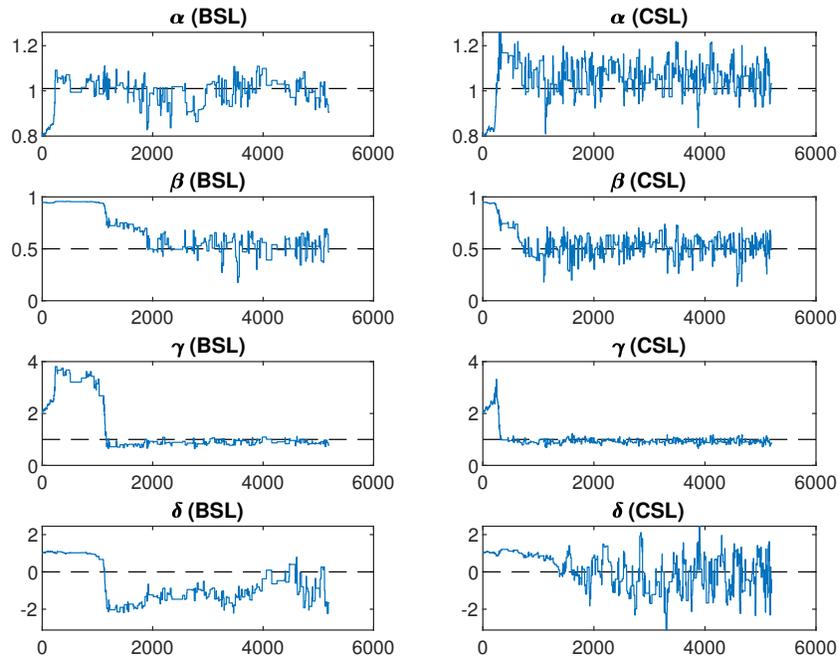}
    \caption{$\alpha$-stable model: (first attempt) trace-plots when using BSL (left) and CSL with G=50 (right).}
    \label{fig:alphastable_traceplots_attempt1}
\end{figure}

\begin{figure}
    \centering
    \includegraphics[scale=0.9]{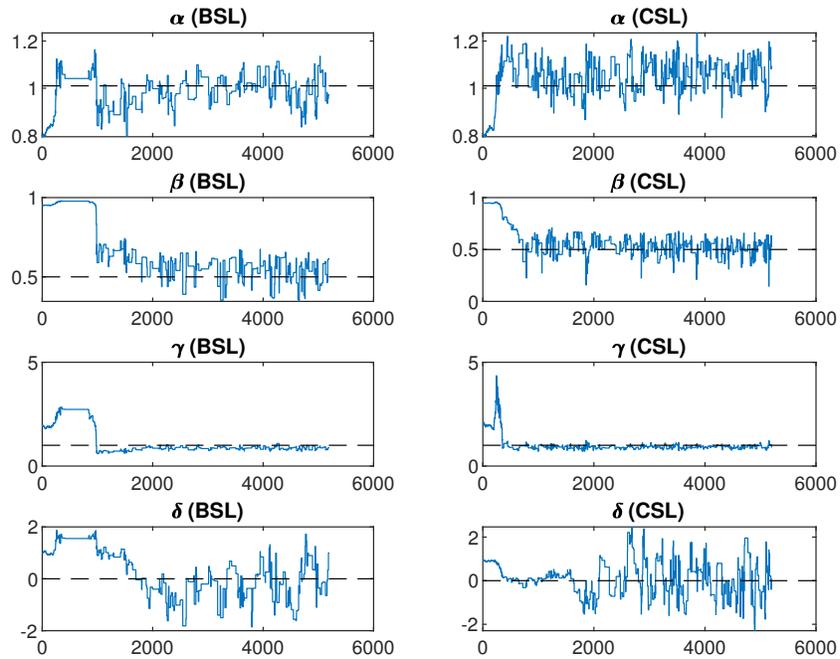}
    \caption{$\alpha$-stable model: (second attempt) trace-plots when using BSL (left) and CSL with G=50 (right).}
    \label{fig:alphastable_traceplots_attempt2}
\end{figure}

\begin{figure}
    \centering
    \includegraphics[scale=0.7]{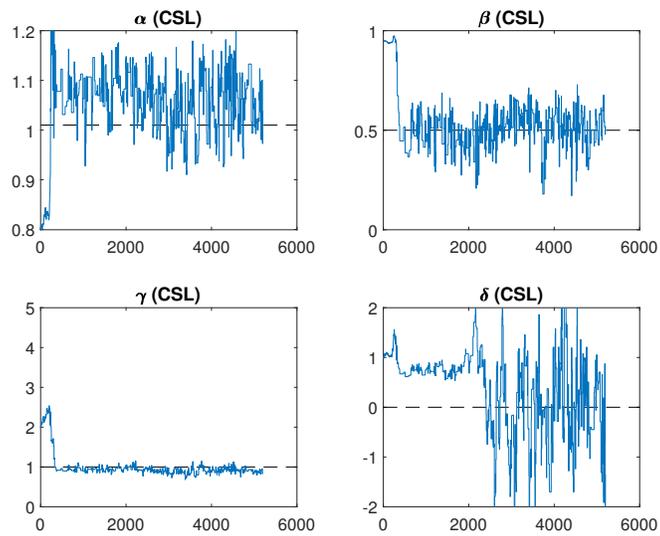}
    \caption{$\alpha$-stable model: (first attempt) trace-plots when using CSL with G=200.}
    \label{fig:alphastable_traceplots_attempt1_G=200}
\end{figure}

\begin{figure}
    \centering
    \includegraphics[scale=0.7]{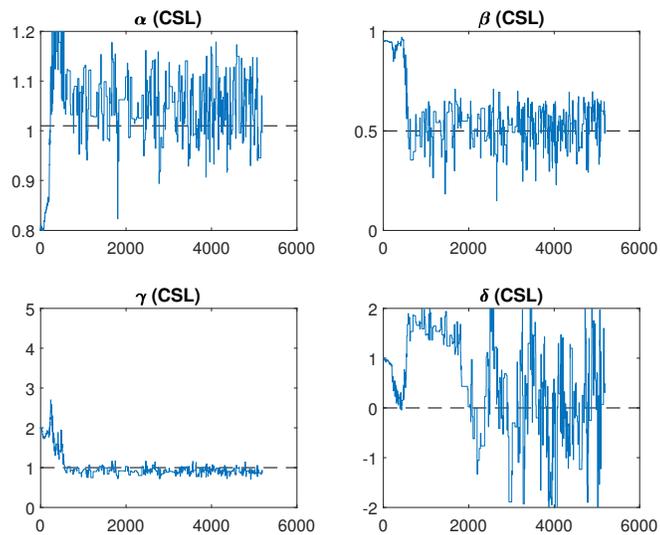}
    \caption{$\alpha$-stable model: (second attempt) trace-plots when using CSL with G=200.}
    \label{fig:alphastable_traceplots_attempt2_G=200}
\end{figure}

Figures \ref{fig:alphastable_traceplots_attempt1}-\ref{fig:alphastable_traceplots_attempt2} report two independent inference attempts, conditionally on the same data, when using BSL and CSL ($G=50$). In all cases the ``Haario'' method is used to produce proposals. Figures \ref{fig:alphastable_traceplots_attempt1_G=200}-\ref{fig:alphastable_traceplots_attempt2_G=200} consider two further attempts for CSL when using $G=200$, always on the same data. The benefits of using an increased value for $G$ (and hence inducing larger correlations) are evident.

\end{document}